\documentclass[useAMS,usenatbib]{mn2e}

\usepackage{mathabx}
\usepackage{booktabs}

\title[UV Rydberg transitions in neutral carbon]
{New atomic data for C I Rydberg states compared with solar UV spectra}
\author[Storey, Dufresne, Del Zanna]{P.~J. Storey$^{1}$, R.P. Dufresne$^{2}$, G. Del Zanna$^{2}$\thanks{Corresponding author E-mail: gd232@cam.ac.uk}
\\
$^{1}$ Department of Physics and Astronomy, University College London, London WC1E 6BT, UK \\
$^{2}$ DAMTP, Centre for Mathematical Sciences, University of Cambridge, Wilberforce Road, Cambridge CB3 0WA, UK \\
}
\date{Submitted to MNRAS  }

\pagerange{\pageref{firstpage}--\pageref{lastpage}} \pubyear{2023}

\DeclareMathAlphabet{\mathsc}{OT1}{cmr}{m}{sc}
\def\testbx{bx}%
\DeclareRobustCommand{\ion}[2]{%
\relax\ifmmode
\ifx\testbx\f@series
{\mathbf{#1\,\mathsc{#2}}}\else
{\mathrm{#1\,\mathsc{#2}}}\fi
\else\textup{#1\,{\mdseries\textsc{#2}}}%
\fi}

\newcommand{\term} [4] {$ {\rm {^#1}{#2^{#3}_{#4}}}$}

\usepackage{graphicx,threeparttable}
\usepackage{txfonts}

\usepackage{epsfig}
\usepackage{color}

\begin{document}

\label{firstpage}
\maketitle

\begin{abstract}
We use the Breit–Pauli $R$-matrix method to calculate accurate energies and radiative data for states in \ion{C}{i} up to $n$=30 and with $l\le 3$. We provide the full dataset of decays to the five 2s$^2$2p$^2$ ground configuration states $^3$P$_{0,1,2}$, $^1$D$_2$, $^1$S$_0$. This is the first complete set of data for  transitions from $n\ge 5$. We compare oscillator strengths and transition probabilities with the few previously calculated values for such transitions, finding generally good agreement (within 10\%) with the exception of values recently recommended by NIST, where significant discrepancies are found. { We then calculate spectral line intensities originating from the Rydberg states using typical chromospheric conditions and assuming LTE,} and compare them with well-calibrated SOHO SUMER UV spectra of the quiet Sun. The relative intensities of the Rydberg series are in excellent agreement with observation, which provides firm evidence for the identifications and blends of nearly 200 UV lines. { Such comparison also resulted in a large number of new identifications of \ion{C}{i} lines in the spectra. We also estimate optical depth effects and find that these can account for much of the absorption noted in the observations.} The atomic data can be applied to model a wide range of solar and astrophysical observations.
\end{abstract}

\begin{keywords}
atomic data --  atomic processes -- Sun: UV radiation -- Sun: chromosphere -- ISM: atoms 
\end{keywords}

\section{Introduction}
\label{sec:intro}

The spectrum of neutral carbon is of importance for a wide range of astrophysical objects and diagnostic applications, across all wavelengths. It has been studied experimentally using laboratory and astrophysical sources for 100 years, since the 1920's. (\citet{haris_kramida:2017ApJS..233...16H} review the observational data, plus a brief overview is given here in Sect.\;\ref{sec:obsoverview}.)

In UV solar spectra of the quiet Sun, there are well over 300 strong spectral lines from neutral carbon that have been observed in the 1100–1700\,\AA\ region. Although the strongest lines in the spectra are resonance and intercombination lines emitted from levels close in energy to the ground, the majority of the \ion{C}{i} lines are emitted from very highly excited, Rydberg levels. Emission has been observed from levels with principal quantum number, $n$, up to 24 in the quiet Sun \citep[][for example]{parenti_etal:2005a} and 29 in flares \citep{feldman_etal:1976_c_1}. Temperatures are too low in the solar chromosphere, where neutral carbon forms, for Rydberg levels to become populated through collisional excitation from the ground configuration of C$^0$. They are, instead, populated through processes linking them to the C$^+$ ground, following photoionisation of C$^0$ due to the solar radiation field \citep{avrett2008}.

With such a wealth of lines comes the opportunity to use them for a variety of purposes. Given the significant Doppler shifts in higher temperature lines in the solar atmosphere, lines from neutrals are often used for instrumental wavelength calibration, unless lines from helium or hydrogen are available. One instance is the Solar Ultraviolet Measurements of Emitted Radiation (SUMER) instrument \citep{wilhelm1995sumer}, on board the Solar and Heliospheric Observatory (SOHO), which observes in 43\,\AA\ spectral windows at a time, and thus requires many lines for calibration across its whole spectral range (660-1610\,\AA\ in the first order). Some uncertainty in the solar wavelengths will be present, and this is reflected in the scatter of values obtained by different instruments and reported in the literature.

The UV region of solar spectra is full of spectral lines from neutral atoms and singly-charged ions, and yet a large fraction are unidentified \citep[see][for example]{sandlin_etal:1986}. Such missing flux is relevant for any diagnostic use of solar UV broad-band images. For example, the Interface Region Imaging Spectrometer (IRIS) provides high-resolution UV broad-band images in two spectral bands. However, they cannot be used for quantitative analyses because they are full of lines for which there is no atomic data. For instance, the current version of the CHIANTI atomic database \citep[v.10][]{dere1997, chianti_v10}, widely used for solar spectroscopic analysis, is rather limited for \ion{C}{i}, containing just 42 levels. 

The UV lines from neutral carbon have also been studied extensively with observations carried out with the Goddard High Resolution Spectrograph on the Hubble Space Telescope, with the aim of using them to measure the carbon abundance in the interstellar medium \citep[ISM, such as][]{welty1999}. As many calculated oscillator strengths were found to disagree significantly, attempts have been made to measure them from the widths of the absorption lines \citep[see for example][and references therein]{federman:2001ApJ...555.1020F}. However, this procedure is limited by the spectral resolution and ability to resolve all the lines. Several decays from C$^0$ states up to $n=6$ were observed, and disagreements between the observations and earlier compilations of atomic data were attributed to the absence of configuration interaction in early theoretical work. 

Modelling the intensities of neutral carbon lines generally requires a complete collisional-radiative model (CRM). Such models need states resolved by total angular momentum, $J$, for the lower $n$, all Rydberg states which contribute to emission either directly or via cascades, plus at least the ground state of C$^+$. Because rates are required for all the relevant atomic processes to connect the levels together, this means the requirement for a large amount of data. Models of these types have been built for the recombination spectrum of neutral hydrogen \citep{Hummer1987} and neutral helium \citep{delzanna_storey:2022}. In the solar chromosphere, radiative transfer and plasma dynamics must also be included for emission from lower levels \citep[such as][for the \ion{C}{i} 1355.85\,\AA\ line]{lin_etal:2017}. However, optical depth effects are not considered important for Rydberg levels because oscillator strengths decrease with increasing $n$ and the line profiles are Gaussian \citep{sandlin_etal:1986}. Rydberg levels are not usually included in chromospheric models because of the requirement for computational speed, although they are included in some cases by grouping them into `superlevels' \citep[such as][]{avrett2008}. The absence of this data causes, for instance, hard continuum edges for neutrals in the synthetic spectrum of hydrostatic, radiative transfer calculations \citep[such as][]{fontenla_etal:2014}, but such edges are not present in the observations because the average emission from the Rydberg levels matches the continuum intensity at the edge.

Accurate atomic data for Rydberg levels are fundamental to our understanding both of neutral carbon itself and the environments in which it emits. The main aim of the present work is to provide such data, and to compare the results with high resolution observations. The focus of the present work is the UV region and the series of decays to the ground configuration (1s$^2$\,2s$^2$\,2p$^2$), which is comprised of the \term{3}{P}{}{0,1,2} levels with series limits of 1101.08\,\AA, 1101.27\,\AA\ and 1101.60\,\AA\ respectively, the \term{1}{D}{}{2} level with a limit at 1240.27\,\AA, and the \term{1}{S}{}{0} level with the series ending at 1445.67\,\AA. (A summary of the existing atomic structure calculations to date is given in Sect.\;\ref{sec:atomicdata}.)

The neutral hydrogen CRM referenced above indicates that levels with $n\approx10$ and higher should be close to local thermal equilibrium (LTE) in the solar chromosphere, which means a large scale model with all the associated processes and rates would not be necessary to model \ion{C}{i} emission from Rydberg levels. Therefore, a secondary aim of the present work is to produce initial models for spectral line intensities in the solar atmosphere. This is a small step towards more complete modelling of the neutral carbon UV spectral range, which is important for the reasons listed above: line identifications, wavelength calibration, calculating missing flux, and modelling solar UV irradiance and \ion{C}{i} UV line emission from lower levels, amongst other things.


The present article is structured as follows: Section\;\ref{sec:obsoverview} summarises some of the main observations of neutral carbon lines in the UV. Section\;\ref{sec:atomicdata} outlines the methods used to calculate energy levels and radiative data, as well as presenting comparisons with literature values. Section\;\ref{sec:comparison} describes the modelling of the Rydberg states to obtain line intensities and presents a comparison between observed and modelled spectra of the quiet Sun in the optically thin case. Selected lines are then modelled including optical depth effects, using a simple escape probability approach, and show improved agreement with observations compared to the optically thin case. The modelling also provides several new line identifications, which are given at the end of the section. Finally, Section\;\ref{sec:conclusions} draws together the conclusions.

\section{Laboratory and solar UV observations of neutral carbon lines}
\label{sec:obsoverview}

Experimental studies of the carbon spectrum in the laboratory  are not trivial, as it is difficult to obtain clean spectra with neutral carbon. Among many earlier studies of the UV radiation, a notable one is the work by \cite{paschen_kruger:1930AnP...399....1P}, where an accurate list of identifications and wavelengths from 1112 to 2583\,\AA\ was produced. This list includes the strongest transitions of the series from $n\le$10 to \term{3}{P}{}{0,1,2}, \term{1}{D}{}{2}, as well as a few decays to \term{1}{S}{}{0} at longer wavelengths, where it appears the sensitivity of the apparatus decreased significantly.

\cite{wilkinson:1955} made accurate measurements of neutral carbon lines with stated uncertainties between 0.001 and 0.005~\AA\ in the 1158--1931\,\AA\ range, using known wavelengths from \ion{Fe}{i}, \ion{Fe}{ii}, and \ion{Cu}{ii} for the calibration. \cite{mazzoni:1981PhyBC.111..379M} were able, with an experimental setup producing absorption lines, to observe the series of decays up to 20d at 1104\,\AA\ down to the ground term. Only decays from $n\ge$4 at 1200\,\AA\ were listed, but with a stated accuracy in the wavelengths of 0.01\,\AA.
The carbon reference wavelengths used by Mazzoni et al. for their calibration were calculated by \cite{johansson:1965}. \cite{johansson:1965} observed visible lines of the series 3s-nl and 3p-nl, with $n$ up to 10. From this, he was able to predict the wavelengths of the 2p-nl series, below 2000\,\AA, with a quoted accuracy of 0.002\,\AA. However, there are some discrepancies with the \cite{wilkinson:1955} values, indicating that the accuracy is probably worse.

Many lines from neutrals were observed with the Naval Research Laboratory (NRL) normal incidence astigmatic spectrograph aboard Skylab during a large solar flare. The instrument resolution {  (full width at half maximum, FWHM)  was about 0.06\,\AA. \cite{feldman_etal:1976_c_1} reported a list} of \ion{C}{i} lines from 1930\,\AA\ down to 1155\,\AA, noting that the sensitivity of the instrument decreased significantly below 1200\,\AA. Decays from up to $n=29$ were observed. \cite{feldman_etal:1976_c_1} also provided a list of predicted wavelengths below 1150\,\AA, down to 1102\,\AA. Many of the observed lines were listed as blended, from known theoretical wavelengths. Wavelengths could be determined with an accuracy of 0.004\,\AA. 
The spectra were calibrated in wavelength using a range of reference values for neutral C, Si, N, and S. The carbon reference wavelengths were those calculated by \cite{johansson:1965}. \citeauthor{feldman_etal:1976_c_1} report that their measurements  seldom deviate by more than 0.01\,\AA\ from the predicted values. 
{ As the experimental energies 
form the database at the National Institute of Standards and Technology \citep[NIST, see][]{NIST_ASD} } are based on all the above compilations, they have an associated uncertainty of about 1~cm$^{-1}$ or more.


A high-resolution solar atlas in the 1175–1700\,\AA\ region of the quiet Sun, limb and an active region were obtained by \cite{sandlin_etal:1986} from the High Resolution Telescope and Spectrometer (HRTS) observations. The HRTS instrument, flown on sounding rockets, had a spectral resolution (0.05\,\AA) similar to that of the Skylab instrument, but produced stigmatic images. Hence, it could resolve more lines and the list contains about 192 neutral carbon lines. As in the previous paper, the wavelength scale was obtained using reference values for several neutrals. That these observations do not include the $^3$P series from Rydberg levels is a limitation for the present study.

A wider UV spectral range (660-1610\,\AA\ in first order) was later observed with the { SOHO SUMER} instrument \citep{wilhelm1995sumer}, although with a lower spectral resolution (about 0.13\,\AA\ FWHM). For the first time, a large number of observations of the quiet Sun was obtained. One limitation of the SUMER instrument mentioned above was that a spectral region of only about 40\,\AA\ could be observed at a time. Another limitation was that at some wavelengths strong second order lines are blending the first order lines. A great asset of the instrument was the radiometric calibration, accurate to within 20\% or so.

\section{Calculation of atomic data}
\label{sec:atomicdata}

\subsection{Overview of existing atomic calculations for neutral carbon Rydberg levels}

There is an extensive literature on radiative data for the lower states of neutral carbon, partly reviewed by \cite{haris_kramida:2017ApJS..233...16H}, but comparatively very little for $n \ge 6$ states. Earlier calculations of radiative data for the lower states of neutral carbon were carried out with, for example, {\sc Superstructure} and the Thomas-Fermi-Amaldi central potential in such works as \cite{nussbaumer_storey:1984}, with the multiconfiguration Hartree–Fock (MCHF) and the Breit–Pauli (MCHF-BP) approximation by, for example, \cite{tachiev:2001CaJPh..79..955T}, or using the CIV3 code (up to $n$=4) with semi-empirical adjustments to the diagonal elements of the Hamiltonian by \cite{hibbert_etal:1993}. Recent accurate calculations for states up to $n=5$ by \cite{li_etal:2021MNRAS.502.3780L}  used the Multiconfiguration Dirac–Hartree–Fock (MCDHF) method, implemented in a parallelized and improved version \citep{jonsson_etal:2013, fischer2019} of the {\sc General-purpose Relativistic Atomic Structure Package} \citep{grant_etal:1980_grasp}.

{ 
Atomic structure calculations such as the above-mentioned ones are known to provide radiative data for Rydberg states that are generally not very accurate.} An alternative approach, the `frozen cores' approximation, was pioneered by M. Seaton in the 1970's, \citep[see for example][]{saraph_seaton:1971RSPTA.271....1S,seaton_wilson:1972JPhB....5L...1S,seaton:1972JPhB....5L..91S}. The idea is to use the framework developed for the scattering calculations for the { $N+1$}  electron system (an { $N$}-electron ion  for the target plus one colliding electron) to calculate the energies and radiative parameters for the bound states. { The wavefunctions of the system are antisymmetrized products of target functions multiplied by the orbital function of the added electron. As part of the Opacity Project (OP) \citep{OPII}, \cite{1985JPhB...18.2111S} described the techniques required to derive bound state energies and radiative data for the $N+1$ electron system within the $R$-matrix formulation of the problem.}

The frozen cores approximation generally produces accurate energies for the Rydberg states  relative to the N-electron system, in this case C$^+$. This is the main reason why this method
was adopted by \cite{oleg:2002} to calculate accurate radiative data for states in neutral carbon up to $n$=10. The authors  adopted the B-splines representation and non-orthogonal one-electron radial orbitals, combined with core states derived from the MCHF method. Only oscillator strengths for decays to the ground levels \term{3}{P}{}{0,1,2} were published, { however}. The only radiative data for $n \ge$10 states available in the literature is for some transition probabilities calculated by \cite{haris_kramida:2017ApJS..233...16H} using the Cowan code \citep{cowan:1981}.

\subsection{ The frozen cores method within the $R$-matrix framework}

The Breit–Pauli $R$-matrix method which is used in this calculation
is described fully elsewhere \citep[see][and references therein]{1993A&A...279..298H,1995CoPhC..92..290B}. 

\begin{table}
\caption{The C$^+$ target configuration basis where the 1s$^2$  core is suppressed. The bar indicates a
correlation orbital.}
\begin{flushleft}
\centering
\begin{tabular}{llll}
\noalign{\hrule}\noalign{\smallskip}
 & 2s$^2$\,2p      & 2s\,2p$^2$ & 2p$^3$  \cr
 & 2s$^2$\,$\overline{3}$l & 2s\,2p\,$\overline{3}$l & 2p$^2$\,$\overline{3}$l \ \ l=0,1,2  \cr
 & 2s\,$\overline{3}$l$^2$  &  2p\,$\overline{3}$l$^2$ \cr
\noalign{\hrule}
\end{tabular}
\end{flushleft}
\label{configs}
\end{table}

\begin{table}
\caption{Potential scaling parameters. The bar over the principal quantum number and the minus sign attached to the value of a scaling parameter signifies a correlation orbital.}
\centering %
\begin{tabular}{lr@{\hskip 1.3cm}lr@{\hskip 1.3cm}lr}
\noalign{\hrule}\noalign{\smallskip}
 1s       & 1.43347 \cr
 2s       & 1.24930 & 2p & 1.21267 \cr
 $\overline{3}$s       & -0.74865 & $\overline{3}$p & -0.70502 & $\overline{3}$d & -0.95464 \cr
 \noalign{\hrule}\noalign{\smallskip}
\end{tabular}
\label{scale}
\end{table}

A set of 18 electron configurations, listed in Table~\ref{configs}, was used to expand the target states. The target wavefunctions were generated with the {\sc Autostructure} program \citep{1974CoPhC...8..270E,1978A&A....64..139N,2011CoPhC.182.1528B} using radial functions computed within scaled Thomas-Fermi-Dirac statistical model potentials. The scaling parameters were determined by minimizing the sum of the energies of all the target terms, computed in { $LS$} coupling, i.e. by neglecting all relativistic effects. The resulting scaling parameters, $\lambda_{nl}$, are given in Table~\ref{scale}. In Table~\ref{termlist} a comparison is made between the term energies calculated using our scattering target and the experimental values for the terms of the ground, $n=2$, complex. The term energies are computed with the inclusion of one-body relativistic effects, the Darwin and mass terms, and the spin–orbit interaction. This is the level of approximation that is available for the scattering calculations in the $R$-matrix code.

\begin{table}
\caption{Energies of the 8 lowest C$^+$ target terms in cm$^{-1}$. The calculated values include only the
spin-orbit contribution to the fine-structure energies. }
\centering %
\begin{tabular}{lllrr}
 \noalign{\hrule}\noalign{\smallskip}
 & & \hspace{1.5cm} & \multicolumn{2}{c}{\hspace{0cm}Term Energy} \cr
 & Config. & Term & Exp.$^{\dagger}$ & Calc.  \cr
 \noalign{\hrule}\noalign{\smallskip}
 & 2s$^2$\,2p  & $^2$P$^{\rm o}$ &     0   &   0   \\
 & 2s\,2p$^2$  & $^4$P          &  42994   & 42073  \\
 &             & $^2$D          &  74889   & 76470  \\
 &             & $^2$S          &  96451  & 99394  \\
 &             & $^3$P          &  110699  & 113418  \\
 & 2p$^3$      & $^4$S$^{\rm o}$  & 142027  & 141729 \\
 &             & $^2$D$^{\rm o}$  & 150463   & 152077  \\
 &             & $^2$P$^{\rm o}$  & 168742  & 174326 \\
  \noalign{\hrule}
 \multicolumn{5}{l}{$^{\dagger}$Experimental energies are from NIST \citep{Kramida&Haris2022} } \\
\end{tabular}
\label{termlist}
\end{table}

A further measure of the quality of the target is a comparison between weighted oscillator strengths, $gf$, calculated in the length and velocity formulations. Good agreement between the two formulations is a necessary but not sufficient condition for ensuring the quality of the target wavefunctions. This comparison is given in Table~\ref{gflv} which also lists the main contributions to the dipole polarisability of the C$^+$ ground state. The Rydberg electron polarises the C$^+$ core electrons and that results in the Rydberg electron experiencing a more attractive potential which lowers its energy below the hydrogenic value. The bulk of the polarisability arises from the terms in the ground complex but there is also a significant contribution (25\%) from the $\overline{3}l$ states where the correlation orbitals provide an approximation to the contributions from all higher target states and the continuum.    

We calculated energies for all the odd parity bound states of neutral carbon up to the energy corresponding to 2s$^2$2p\,($^2$P$^{\rm o}_{1/2}$)\,$nl$, with $n=30$ and $l\le 3$. { The experimental energy difference between the C$^+$ $^2$P$^{\rm o}_{1/2}$ and $^2$P$^{\rm o}_{3/2}$ levels is 63.4\,cm$^{-1}$ and the calculated value of 62.8\,cm$^{-1}$ was corrected to the experimental value before the Rydberg level energies were calculated. The $n=30$ levels in the $^2$P$^{\rm o}_{1/2}\,nl$ series then correspond to $n=24$ in the $^2$P$^{\rm o}_{3/2}\,nl$ series. } 

\begin{figure}
\centering
\includegraphics[width=1.0\columnwidth]{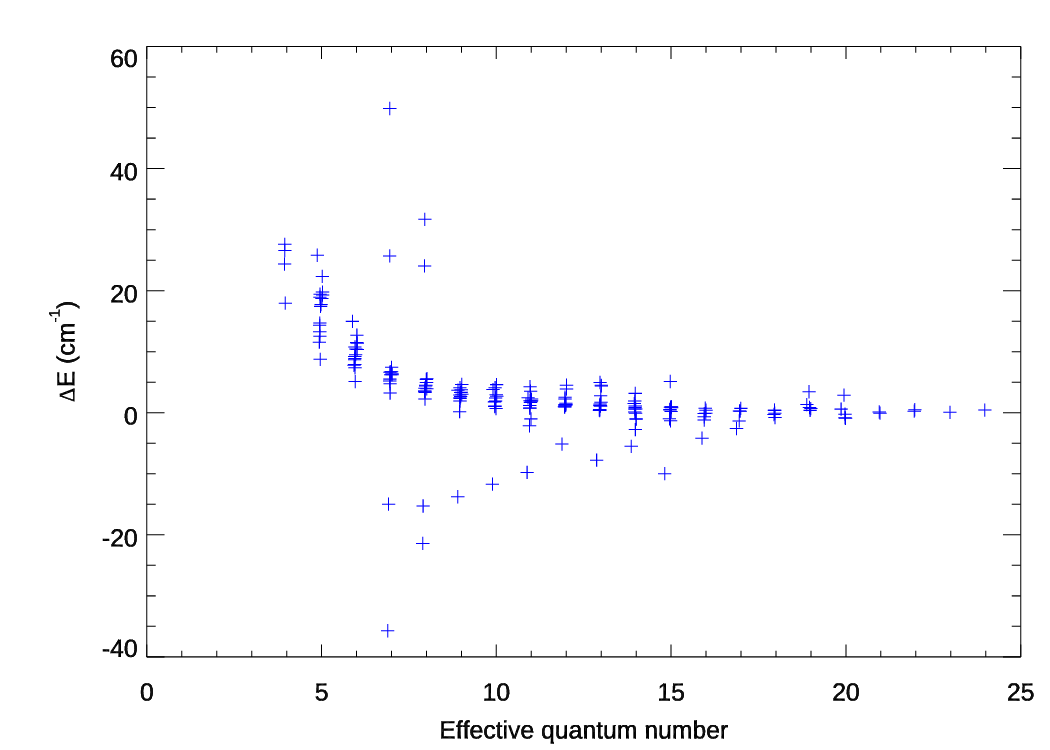}
\caption{Energy differences against effective quantum number relative to the C$^+$ ground level between the present calculation and all experimentally known states with $n\ge5$ and $l\le3$.}
\label{fig:energies_vs_nist}
\end{figure}

Figure~\ref{fig:energies_vs_nist} shows the difference between calculated and experimental energies where experimental values are available. Experimental energies are all taken from NIST \citep{NIST_ASD}. Since this calculation was designed for Rydberg states we compare our results only for $n\ge 5$, and the agreement with experiment improves considerably with increasing $n$, as expected. An energy difference of 1\,cm$^{-1}$ corresponds to a difference in transition wavelength of 12\,m\AA\ at 1101\,\AA. With a few exceptions, all states with $n\ge 15$ are within 1\,cm$^{-1}$ of experiment, giving us confidence that the wavelengths of transitions from high Rydberg states which are not experimentally known can be predicted with an accuracy comparable to experimental methods.  { This high accuracy can be achieved in part because the energies are calculated relative to the C$^+$ target ground state, which is experimentally known relative to the C$^0$ ground state. } The group of states with systematically negative and relatively larger differences are those with an $n$s orbital, which are more difficult to calculate due to the penetration of the s orbitals to small radial distances.

\begin{table}
\caption{Weighted $LS$ oscillator strengths, $gf$, in the length and velocity formulations from the C$^+$ target
ground state and the main contributions to the ground state dipole polarisability, $\alpha_{\rm D}$ in atomic units.} \centering
\begin{tabular}{lllllrr@{\hskip 1.5cm}rr}
\noalign{\hrule}
 & \multicolumn{5}{c}{\hspace{-1.5cm}Transition}  & $gf_{_L}$ & $gf_{_V}$ &  $\alpha_{\rm D}$ \\
 \noalign{\hrule}\noalign{\smallskip}
 & 2s$^2$\,2p & $^2$P$^{\rm o}$  & -- & 2s\,2p$^2$ & $^2$D  & 0.84  &  0.86 & 1.16      \\
 &            &                & -- &            & $^2$S  & 0.67  & 0.75 & 0.55  \\
 &            &                & -- &            & $^2$P    & 3.11  & 3.11 & 1.94  \\
 &            &                & -- & 2s$^2$\,$\overline{3}$s & $^2$S & 0.33 & 0.26 & 0.14 \\
 &           &                 & -- & 2s\,2p\,$\overline{3}$p & $^2$P & 2.75 & 2.99 & 0.40 \\
 &                     &       & -- & 2s$^2$\,$\overline{3}$d & $^2$D & 6.25 & 4.71 & 0.68 \\
\noalign{\hrule}
\end{tabular}
\label{gflv}
\end{table}

We obtained oscillator strengths from these states to the five levels of the 2s$^2$2p$^2$ ground configuration, $^3$P$_{0,1,2}$, $^1$D$_2$, $^1$S$_0$. In Table~\ref{fcomp1} we compare our results for the transitions from upper $n$d states with $J=3^{\rm o}$ to the lower $^3$P$_2$ with those of \citet[][ZF]{oleg:2002} and \citet[][HK]{haris_kramida:2017ApJS..233...16H} for $5\le n \le 9$. These transitions give rise to many of the strongest lines in the solar spectrum. \cite{haris_kramida:2017ApJS..233...16H} quote the results of the calculation by \cite{tachiev:2001CaJPh..79..955T} for $n\le 8$ and for these transitions the agreement is excellent. For $n=9$ and above they quote the results of their own calculations using the code of \cite{cowan:1981} and for $n=9$ the agreement is poor. { We note that \cite{haris_kramida:2017ApJS..233...16H} cite an uncertainty of 50\% for these calculations.} We agree well for all transitions with \cite{oleg:2002}, who also use a close-coupling method well suited to the treatment of Rydberg states. We will return to the comparison with \cite{haris_kramida:2017ApJS..233...16H} for $n\ge9$ below.

\begin{table}
  \caption{Comparison of absorption oscillator strengths in the length formulation for 2s$^2$\,2p$^2$\,$^3$P$_2$  -- 2s$^2$\,2p\,$n$d ($J=3^{\rm o}$)
    transitions with $5\le n \le 9$ from {\citet{oleg:2002}}, (ZF),
    the compilation of {\citet{haris_kramida:2017ApJS..233...16H}}, (HK) and the present work (PW). For $n \le 8$,
    {\citet{haris_kramida:2017ApJS..233...16H}} quote the calculations of {\citet{tachiev:2001CaJPh..79..955T}}.
    For $n$=9 and higher they report their own calculations using the code described by {\citet{cowan:1981}}.}
  \centering
\begin{tabular}{lllllll}
\hline
\noalign{\hrule}
 Level & $\lambda$[\AA] & $f_L$ & $f_L$  & $f_L$ \\
        &    &  ZF   &   HK  &   PW  \\
            \hline
 2s$^2$2p\,($^2$P$_{1/2}^{\rm o}$) 5d [5/2] &  1159.0 & 2.04(-3) &  2.03(-3) & 1.90(-3) \\ 
 2s$^2$2p\,($^2$P$_{3/2}^{\rm o}$) 5d [5/2] &  1158.0 & 1.57(-2) & 1.58(-2) & 1.59(-2)  \\ 
 2s$^2$2p\,($^2$P$_{3/2}^{\rm o}$) 5d [7/2] &   1157.4 & 1.12(-3) & 1.12(-3) & 1.12(-3) \\ 
 2s$^2$2p\,($^2$P$_{1/2}^{\rm o}$) 6d [5/2] &  1140.7 & 1.72(-3) & 1.72(-3) & 1.67(-3) \\ 
 2s$^2$2p\,($^2$P$_{3/2}^{\rm o}$) 6d [5/2] &  1139.9 & 8.07(-3) & 8.18(-3) & 8.21(-3) \\ 
 2s$^2$2p\,($^2$P$_{3/2}^{\rm o}$) 6d [7/2] &  1139.5 & 1.17(-3) & 1.17(-3) & 1.18(-3) \\ 
 2s$^2$2p\,($^2$P$_{1/2}^{\rm o}$) 7d [5/2] &  1130.0 & 1.30(-3) & 1.29(-3) & 1.29(-3) \\ 
 2s$^2$2p\,($^2$P$_{3/2}^{\rm o}$) 7d [5/2] &  1129.2 & 4.56(-3) & 4.55(-3) & 4.63(-3) \\ 
 2s$^2$2p\,($^2$P$_{3/2}^{\rm o}$) 7d [7/2] &  1129.0 & 1.06(-3) & 1.07(-3) & 1.07(-3)  \\ 
 2s$^2$2p\,($^2$P$_{1/2}^{\rm o}$) 8d [5/2] &  1123.2 & 9.47(-4) & 9.53(-4) & 9.53(-4) \\ 
 2s$^2$2p\,($^2$P$_{3/2}^{\rm o}$) 8d [5/2] &  1122.5 & 2.80(-3) & 2.80(-3) & 2.84(-3) \\ 
 2s$^2$2p\,($^2$P$_{3/2}^{\rm o}$) 8d [7/2] &  1122.3 & 8.93(-4) & 8.99(-4) & 8.96(-4) \\ 
 2s$^2$2p\,($^2$P$_{1/2}^{\rm o}$) 9d [5/2] &  1118.6 & 6.93(-4) & 1.05(-3) & 7.01(-4)  \\ 
 2s$^2$2p\,($^2$P$_{3/2}^{\rm o}$) 9d [5/2] &  1117.9 & 1.84(-3) & 3.15(-3) & 1.86(-3) \\ 
 2s$^2$2p\,($^2$P$_{3/2}^{\rm o}$) 9d [7/2] &  1117.8 & 7.27(-4) & 6.03(-4)  & 7.30(-4) \\ 
\hline
\end{tabular}
\label{fcomp1}
\end{table}

\begin{table}
\caption{Comparison of absorption oscillator strengths in the length formulation for 2s$^2$\,2p$^2$\,$^1$D$_2$ -- 2s$^2$\,2p\,$n$d ($J=3^{\rm o}$) transitions with $4\le n \le 9$ from \citet{li_etal:2021MNRAS.502.3780L}, (Li21), the $LS$-coupling results of \citet{1989JPhB...22.3377L} (L89), the compilation of \citet{haris_kramida:2017ApJS..233...16H} (HK) and the present work (PW). For $n=4$, \citet{haris_kramida:2017ApJS..233...16H} quote the calculations of \citet{hibbert_etal:1993}. For $n\ge4$ they report either the $LS$-coupling results of \citet{1989JPhB...22.3377L}, or their own calculations using the code described by \citet{cowan:1981}.} 
\centering
\setlength\tabcolsep{4pt}
\begin{tabular}{lllllll}
\hline
\noalign{\hrule}
 Level & $\lambda$[\AA] & $f_{\rm L}$ & $f_{\rm L}$  & $f_{\rm L}$ & $f_{\rm L}$ \\
        &    &  Li21   &  L89 &   HK  & PW  \\
            \hline
 2s$^2$2p\,($^2$P$_{1/2}^{\rm o}$) 4d $^3$F$^{\rm o}_3$ &  1359.3 & 8.88(-4) &           &  8.53(-4) & 8.15(-4) \\ 
 2s$^2$2p\,($^2$P$_{3/2}^{\rm o}$) 4d $^3$D$^{\rm o}_3$ &  1357.7 & 4.61(-4) &           &  4.26(-4) & 6.34(-4)  \\ 
 2s$^2$2p\,($^2$P$_{3/2}^{\rm o}$) 4d $^1$F$^{\rm o}_3$ &  1355.9 & 3.74(-2) &           &  4.01(-2) & 3.58(-2) \\
   Sum                                                &      & 3.87(-2) &  4.01(-2) &  4.14(-2) & 3.72(-2) \\
 2s$^2$2p\,($^2$P$_{1/2}^{\rm o}$) 5d $^3$F$^{\rm o}_3$ &  1313.5 & 1.22(-3) &           &  2.2(-3)  &  1.14(-3) \\ 
 2s$^2$2p\,($^2$P$_{3/2}^{\rm o}$) 5d $^3$D$^{\rm o}_3$ &  1312.3 & 4.82(-4) &           &  1.5(-4)  & 7.15(-4)  \\ 
 2s$^2$2p\,($^2$P$_{3/2}^{\rm o}$) 5d $^1$F$^{\rm o}_3$ &  1311.4 &          &           &  2.13(-2) & 1.74(-2) \\ 
   Sum                                             &         &          &  2.13(-2) &  2.37(-2) & 1.93(-2) \\
 2s$^2$2p\,($^2$P$_{1/2}^{\rm o}$) 6d $^3$F$^{\rm o}_3$ &  1290.0 &          &           &  2.45(-3) & 1.28(-3) \\ 
 2s$^2$2p\,($^2$P$_{3/2}^{\rm o}$) 6d $^3$D$^{\rm o}_3$ &  1288.9 &          &           &  7.3(-5)  & 5.45(-4)  \\ 
 2s$^2$2p\,($^2$P$_{3/2}^{\rm o}$) 6d $^1$F$^{\rm o}_3$ &  1288.4 &          &           &  1.22(-2) & 9.37(-3) \\
   Sum                                             &         &          &  1.22(-2) &  1.47(-2) & 1.12(-2) \\
 2s$^2$2p\,($^2$P$_{1/2}^{\rm o}$) 7d [5/2]           &  1276.3 &          &           &  2.39(-3) & 1.21(-3) \\ 
 2s$^2$2p\,($^2$P$_{3/2}^{\rm o}$) 7d [5/2]           &  1275.3 &          &           &           & 4.04(-4) \\ 
 2s$^2$2p\,($^2$P$_{3/2}^{\rm o}$) 7d [7/2]           &  1275.0 &          &           &  7.68(-3) & 5.48(-3)  \\
   Sum                                             &         &          &  7.68(-3) &           & 7.09(-3) \\ 
 2s$^2$2p\,($^2$P$_{1/2}^{\rm o}$) 8d [5/2]           &  1267.6 &          &           &  2.4(-3)  & 1.03(-3) \\ 
 2s$^2$2p\,($^2$P$_{3/2}^{\rm o}$) 8d [5/2]           &  1266.6 &          &           &           & 2.87(-4) \\ 
 2s$^2$2p\,($^2$P$_{3/2}^{\rm o}$) 8d [7/2]           &  1266.4 &          &           &  5.12(-3) & 3.44(-3) \\ 
   Sum                                             &         &          &  5.12(-3) &            & 4.78(-3) \\
 2s$^2$2p\,($^2$P$_{1/2}^{\rm o}$) 9d [5/2]           &  1261.7 &          &           &  2.(-3) & 8.49(-4)  \\ 
 2s$^2$2p\,($^2$P$_{3/2}^{\rm o}$) 9d [5/2]           &  1260.7 &          &           &           & 2.05(-4) \\ 
 2s$^2$2p\,($^2$P$_{3/2}^{\rm o}$) 9d [7/2]           &  1260.6 &          &           &   6.0(-3) & 2.29(-3) \\
  Sum                                                 &         &          &  3.66(-3) &           &3.34(-3) \\

 \hline
\end{tabular}
\label{fcomp2}
\end{table}

\begin{figure}
\centering
\includegraphics[width=1.0\columnwidth]{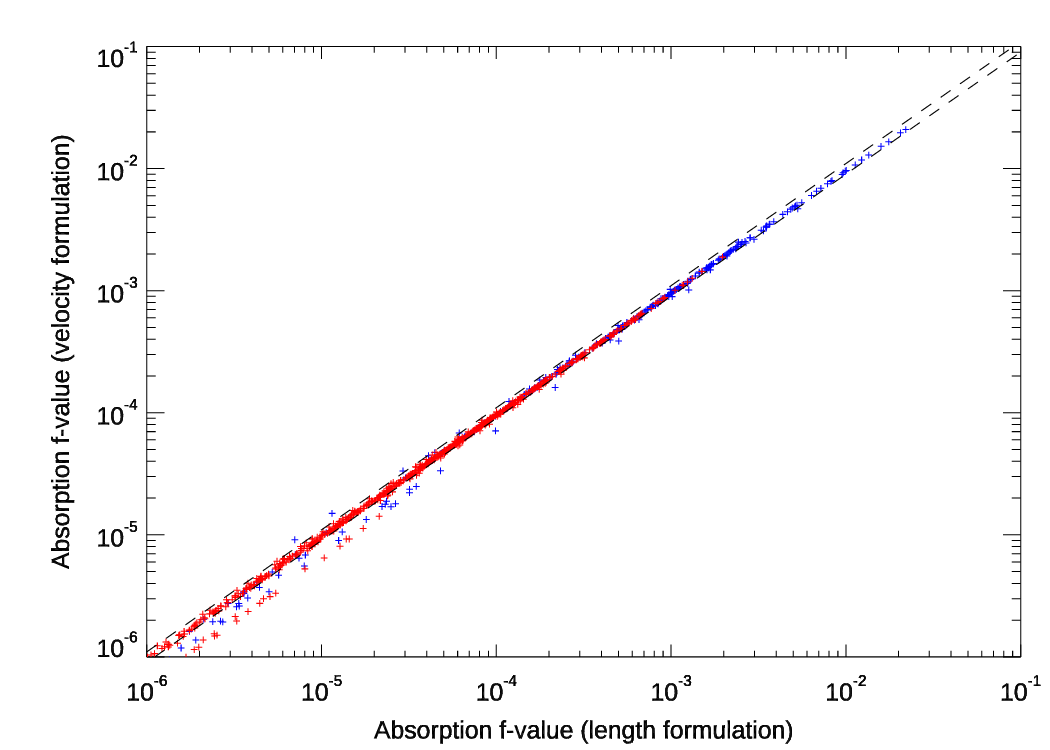}
\caption{ Comparison of oscillator strengths from the velocity formulation with those from the length formulation. Blue crosses represent oscillator strengths for transitions with upper levels in the range $5\le n\le 10$ and red crosses are those from levels with $n\ge 11$. The black dashed lines represent differences of $\pm 10\%$.}
\label{fig:f_values}
\end{figure}

In Table \ref{fcomp2} we make a similar comparison for transitions from the upper $n$d states with $J=3^{\rm o}$ to the lower $^1$D$_2$ level. In this table we have retained the NIST labelling convention of the states in that for $n\le 6$, $LSJ$-coupling notation is used while higher states are described by a pair-coupling notation. The transition between coupling schemes is evident in the behaviour of the oscillator strengths in that for low $n$, the spin changing transitions are much weaker but grow in strength as $n$ increases and the fine-structure interactions become stronger. \citet[][L21]{li_etal:2021MNRAS.502.3780L} reported the results { of} an MCDHF calculation for states with $n\le 4$ and two of the $n=5$ states. We find good agreement with our work for the strong transition from the 4d\,$^1$F$_3$ state but less good for the weaker spin-changing transitions. A similar picture emerges when comparing with the compilation of \cite{haris_kramida:2017ApJS..233...16H}, who quote the work of \cite{hibbert_etal:1993} for the transitions from the $n=4$ states. 

We can also compare with the results from the Opacity Project  \citep{1987JPhB...20.6363S, OPII} reported by \citet[][L89]{1989JPhB...22.3377L}
who used the same techniques as the present work but in $LS$-coupling, quoting oscillator strengths for the $n$d\,$^1$F$^{\rm o}$  -- $^1$D transitions. As $n$ increases, the oscillator strength becomes distributed among the three $n$d $J=3$ levels, so it is appropriate to compare our total oscillator strength for a given $n$ to the values calculated by   \cite{1989JPhB...22.3377L}. The agreement is excellent, within 10\% or less. \cite{haris_kramida:2017ApJS..233...16H} also cite the OP results but attribute the whole of the  $n$d\,$^1$F$^{\rm o}$  -- $^1$D oscillator strength to the $n$d\,$^1$F$^{\rm o}_3$  -- $^1$D$_2$ transition which overestimates that oscillator strength increasingly as $n$ increases and the spin-changing transitions take up more of the oscillator strength. { \cite{haris_kramida:2017ApJS..233...16H} cite the work of \cite{WieseFuhr1996} and \cite{WieseFuhr2007} as the source of these values in their compilation.} As mentioned above, for $n > 8$ \cite{haris_kramida:2017ApJS..233...16H} quote the results of their own calculation using the code of \cite{cowan:1981} and we again find large differences from their work.  

In Table~\ref{Rydstates} we list the calculated energies of states with $n\ge 10$ and $J=3$, odd parity, which give rise to some of the strongest observed transitions.   Table~\ref{Rydstates} lists the transition probabilities from each of the states to the 2s$^2$2p$^2$\, $^3$P$_2$ state from our calculation and from the compilation of \cite{haris_kramida:2017ApJS..233...16H}, which is the only other source of which we are aware for transition probabilities for some of the high $n$ Rydberg states. The good agreement that we find for lower $n$ transitions between our results and those of \cite{oleg:2002}, \cite{tachiev:2001CaJPh..79..955T} for the $n$d -- $^3$P$_2$ transitions and \cite{1989JPhB...22.3377L} for the $n$d -- $^1$D$_2$ transitions leads us to prefer our results over those of \cite{haris_kramida:2017ApJS..233...16H} for the higher Rydberg states, when they differ significantly as they do in Table~\ref{Rydstates}.

{ As discussed above in relation to the scattering target, good agreement between the length and velocity formulations of the oscillator strengths is a necessary but not sufficient condition for ensuring accuracy (see also \citealt{Kramida2013}). For the full set of oscillator strengths reported here the average absolute difference between the length and velocity forms is within 10\%. For the stronger transitions with absorption oscillator strengths larger than 10$^{-5}$ the difference is only 5\%. This is illustrated in Fig.\;\ref{fig:f_values}, where it can be seen that good agreement is maintained for transitions from states from both high and low $n$. There is one anomalous series, for weaker transitions from $J=2$ states to $2s^2\,2p^2\;^1D_2$, where differences reach 70\%. However, the velocity form tends to be much more variable than the length form as a function of basis size and quality, so that the uncertainty in the oscillator strength is usually smaller than the difference between the two formulations. \cite{Kramida2013} also suggests examining the behaviour of series of oscillator strengths from a given lower level for signs of irregularity. This is not suitable in this case due to the overlapping and resulting interaction between levels of different $n$ converging on the two ground levels of C$^+$. }

The complete list of all lines produced for this work has been made available at ZENODO { (DOI: 10.5281/zenodo.8225753}. 
The line list is ordered by calculated wavelength and includes theoretical and observed wavelengths, where available, absorption oscillator strength { in both the length and velocity formulation,} transition decay probability, upper and lower configurations, as well as a measure of the emissivity in the line. A sample of this table is shown in Table\;\ref{tab:datasample}.

\begin{table}
\caption{The C$^0$ odd parity $n$d Rydberg states with total $J=3$.  The calculated energies $E_{\rm Calc}$ are relative to the ground state in Rydbergs. $\Delta\,E$ is the energy difference between calculation and experiment in cm$^{-1}$. $\lambda$ is the wavelength of the transition from this state to the 2s$^2$\,2p$^2$\,$^3$P$_2$ state and $A$ is the corresponding transition probability from this work (PW) and from the compilation of \citet{haris_kramida:2017ApJS..233...16H} (HK). } 
\footnotesize
\setlength\tabcolsep{4pt}
\begin{tabular}{llllll}
\hline\noalign{\hrule}
 Level & $E_{\rm Calc}$ & $\Delta\,E$ & $\lambda$[\AA] & $A$[s$^{-1}$] & $A$[s$^{-1}$] \\
            &     &        &  & PW &  HK \\
            \hline
 2s$^2$2p\,($^2$P$_{1/2}^{\rm o}$) 10d [5/2] &  0.8175273 &    1.9 &    1115.23 &   2.00(+6) & 3.3(+6)\\ 
 2s$^2$2p\,($^2$P$_{3/2}^{\rm o}$) 10d [5/2] &  0.8180900 &    1.8 &    1114.46 &   4.93(+6) & 9.0(+6) \\
 2s$^2$2p\,($^2$P$_{3/2}^{\rm o}$) 10d [7/2] &  0.8181491 &    2.4 &    1114.38 &   2.28(+6) & 1.9(+6) \\ 
 2s$^2$2p\,($^2$P$_{1/2}^{\rm o}$) 11d [5/2] &  0.8192858 &    1.2 &    1112.82 &   1.53(+6) & 2.6(+6) \\ 
 2s$^2$2p\,($^2$P$_{3/2}^{\rm o}$) 11d [5/2] &  0.8198505 &   -2.1 &    1112.01 &   3.54(+6) & 7.0(+6) \\ 
 2s$^2$2p\,($^2$P$_{3/2}^{\rm o}$) 11d [7/2] &  0.8198949 &    1.7 &    1112.00 &   1.87(+6) & 1.7(+6)\\ 
 2s$^2$2p\,($^2$P$_{1/2}^{\rm o}$) 12d [5/2] &  0.8206217 &    1.3 &    1111.01 &   1.19(+6) & 2.2(+6) \\ 
 2s$^2$2p\,($^2$P$_{3/2}^{\rm o}$) 12d [5/2] &  0.8211880 &    1.0 &    1110.24 &   2.60(+6) & 6.0(+6) \\ 
 2s$^2$2p\,($^2$P$_{3/2}^{\rm o}$) 12d [7/2] &  0.8212222 &    1.2 &    1110.20 &   1.58(+6) & 1.6(+6) \\ 
 2s$^2$2p\,($^2$P$_{1/2}^{\rm o}$) 13d [5/2] &  0.8216603 &    1.2 &    1109.61 &   9.40(+5) & 3.6(+6) \\ 
 2s$^2$2p\,($^2$P$_{3/2}^{\rm o}$) 13d [5/2] &  0.8222279 &    0.4 &    1108.83 &   1.91(+6) & 9.0(+6) \\ 
 2s$^2$2p\,($^2$P$_{3/2}^{\rm o}$) 13d [7/2] &  0.8222545 &    1.2 &    1108.80 &   1.39(+6) & 2.9(+6) \\ 
 2s$^2$2p\,($^2$P$_{1/2}^{\rm o}$) 14d [5/2] &  0.8224840 &    0.6 &    1108.49 &   7.47(+5) & 1.3(+6) \\ 
 2s$^2$2p\,($^2$P$_{3/2}^{\rm o}$) 14d [5/2] &  0.8230517 &    1.5 &    1107.73 &   1.23(+6) & 2.3(+6) \\ 
 2s$^2$2p\,($^2$P$_{3/2}^{\rm o}$) 14d [7/2] &  0.8230725 &    1.1 &    1107.70 &   1.47(+6) & 3.3(+6) \\ 
 2s$^2$2p\,($^2$P$_{1/2}^{\rm o}$) 15d [5/2] &  0.8231495 &    0.5 &    1107.59 &   5.45(+5) & 2.6(+5) \\ 
 2s$^2$2p\,($^2$P$_{1/2}^{\rm o}$) 16d [5/2] &  0.8236838 &   -0.1 &    1106.86 &   3.20(+5) \\ 
 2s$^2$2p\,($^2$P$_{3/2}^{\rm o}$) 15d [5/2] &  0.8237200 &   -1.0 &    1106.80 &   2.00(+6) & 3.0(+6)\\ 
 2s$^2$2p\,($^2$P$_{3/2}^{\rm o}$) 15d [7/2] &  0.8237382 &    0.9 &    1106.80 &   3.32(+5) \\ 
 2s$^2$2p\,($^2$P$_{1/2}^{\rm o}$) 17d [5/2] &  0.8241373 &    0.3 &    1106.26 &   4.12(+5) & 7.0(+5) \\ 
 2s$^2$2p\,($^2$P$_{3/2}^{\rm o}$) 16d [5/2] &  0.8242619 &   -0.7 &    1106.08 &   1.21(+6) \\ 
 2s$^2$2p\,($^2$P$_{3/2}^{\rm o}$) 16d [7/2] &  0.8242765 &    0.7 &    1106.08 &   5.77(+5) & 2.2(+6) \\ 
 2s$^2$2p\,($^2$P$_{1/2}^{\rm o}$) 18d [5/2] &  0.8245146 &    0.4 &    1105.75 &   3.56(+5) \\ 
 2s$^2$2p\,($^2$P$_{3/2}^{\rm o}$) 17d [5/2] &  0.8247118 &   -1.3 &    1105.47 &   8.87(+5) & 1.7(+6) \\ 
 2s$^2$2p\,($^2$P$_{3/2}^{\rm o}$) 17d [7/2] &  0.8247239 &    0.3 &    1105.47 &   5.94(+5) \\ 
 2s$^2$2p\,($^2$P$_{1/2}^{\rm o}$) 19d [5/2] &  0.8248338 &    0.9 &    1105.33 &   3.01(+5) \\ 
 2s$^2$2p\,($^2$P$_{3/2}^{\rm o}$) 18d [5/2] &  0.8250875 &   -0.3 &    1104.98 &   3.05(+5) & 1.7(+6)\\ 
 2s$^2$2p\,($^2$P$_{3/2}^{\rm o}$) 18d [7/2] &  0.8250963 &   -0.1 &    1104.97 &   1.12(+6) \\ 
 2s$^2$2p\,($^2$P$_{1/2}^{\rm o}$) 20d [5/2] &  0.8251099 &   -0.2 &    1104.95 &   7.64(+4) \\ 
 2s$^2$2p\,($^2$P$_{1/2}^{\rm o}$) 21d [5/2] &  0.8253385 &    0.1 &    1104.64 &   2.18(+5) \\ 
 2s$^2$2p\,($^2$P$_{3/2}^{\rm o}$) 19d [5/2] &  0.8254084 &    3.4 &    1104.59 &   7.15(+5) & 1.3(+6) \\ 
 2s$^2$2p\,($^2$P$_{3/2}^{\rm o}$) 19d [7/2] &  0.8254171 &    0.4 &    1104.54 &   3.52(+5) & 2.3(+5) \\ 
 2s$^2$2p\,($^2$P$_{1/2}^{\rm o}$) 22d [5/2] &  0.8255417 &    0.3 &    1104.37 &   1.96(+5) \\ 
 2s$^2$2p\,($^2$P$_{3/2}^{\rm o}$) 20d [5/2] &  0.8256803 &    2.9 &    1104.22 &   4.81(+5) & 1.1(+6) \\ 
 2s$^2$2p\,($^2$P$_{3/2}^{\rm o}$) 20d [7/2] &  0.8256875 &   -0.8 &    1104.17 &   4.41(+5) & 2.6(+5) \\ 
 2s$^2$2p\,($^2$P$_{1/2}^{\rm o}$) 23d [5/2] &  0.8257193 & &    1104.13 &   1.59(+5) \\ 
 2s$^2$2p\,($^2$P$_{1/2}^{\rm o}$) 24d [5/2] &  0.8258731 & &    1103.93 &   1.44(+5) \\ 
 2s$^2$2p\,($^2$P$_{3/2}^{\rm o}$) 21d [5/2] &  0.8259149 & &    1103.87 &   5.40(+5) \\ 
 2s$^2$2p\,($^2$P$_{3/2}^{\rm o}$) 21d [7/2] &  0.8259214 &   -0.1 &    1103.86 &   2.52(+5) & 2.5(+5) \\ 
 2s$^2$2p\,($^2$P$_{1/2}^{\rm o}$) 25d [5/2] &  0.8260104 & &    1103.74 &   1.33(+5) \\ 
 2s$^2$2p\,($^2$P$_{3/2}^{\rm o}$) 22d [5/2] &  0.8261173 & &    1103.60 &   2.84(+5) \\ 
 2s$^2$2p\,($^2$P$_{3/2}^{\rm o}$) 22d [7/2] &  0.8261225 &    0.5 &    1103.60 &   4.42(+5) \\ 
 2s$^2$2p\,($^2$P$_{1/2}^{\rm o}$) 26d [5/2] &  0.8261330 & &    1103.58 &   7.65(+4) \\ 
 2s$^2$2p\,($^2$P$_{1/2}^{\rm o}$) 27d [5/2] &  0.8262396 & &    1103.44 &   1.05(+5) \\ 
 2s$^2$2p\,($^2$P$_{3/2}^{\rm o}$) 23d [5/2] &  0.8262947 & &    1103.36 &   3.66(+5) \\ 
 2s$^2$2p\,($^2$P$_{3/2}^{\rm o}$) 23d [7/2] &  0.8262996 &    0.1 &    1103.36 &   2.35(+5) \\ 
 2s$^2$2p\,($^2$P$_{1/2}^{\rm o}$) 28d [5/2] &  0.8263367 & &    1103.31 &   9.38(+4) \\ 
 2s$^2$2p\,($^2$P$_{1/2}^{\rm o}$) 29d [5/2] &  0.8264230 & &    1103.19 &   8.15(+4) \\ 
 2s$^2$2p\,($^2$P$_{3/2}^{\rm o}$) 24d [5/2] &  0.8264501 & &    1103.16 &   3.54(+5) \\ 
 2s$^2$2p\,($^2$P$_{3/2}^{\rm o}$) 24d [7/2] &  0.8264545 &    0.4 &    1103.16 &   1.77(+5) \\ 
 2s$^2$2p\,($^2$P$_{1/2}^{\rm o}$) 30d [5/2] &  0.8265017 & &    1103.09 &   7.72(+4) \\ 
 \hline
\end{tabular}
\label{Rydstates}
\normalsize
\end{table}

\section{Comparing the atomic data with solar observations}
\label{sec:comparison}



The atomic data { have been assessed further by comparing them with observations. Recent laboratory experiments on} neutral carbon are few and far between, for the reasons discussed in Sect.\;\ref{sec:obsoverview}, but there are numerous observations of neutral carbon emission from Rydberg states in the solar atmosphere.

\subsection{Source of solar observations}

The Skylab flare list includes more lines than the quiet Sun SUMER lists, but intensities were not provided by \citet{feldman_etal:1976_c_1}. Flare spectra are also expected to be more complex to model. As the aim here is to show a comparison between LTE relative intensities and well-calibrated solar radiances, the SUMER quiet Sun spectra have been chosen. \cite{curdt_etal:2001} provided a complete line list of wavelengths from 680 to 1611\,\AA, merging observations obtained over a time span of several hours on 1997-04-20. 



\cite{parenti_etal:2005a} also published a list of wavelengths in the 800-1250\,\AA\ SUMER range, using measured intensities for a quiet Sun observation of 1999-10-09. We have processed the data related to the 1999-10-09 observation to include spectra at longer wavelengths, up to 1322\,\AA. We have used the level 1 calibrated data, and considered only the central part of the detector A, spanning 20\,\AA. The exposure time was 200 s, and each slit exposure was taken about every 4 minutes. Examining the overlapping regions within these observations (of about 8\,\AA), it was possible to assess that very little variability between exposures (a few percent at most) was present in the lines from neutrals. Therefore, we can safely compare the SUMER intensities of lines at very different wavelengths, unlike transitions formed at higher temperatures, where significant variability is observed.

Further, we have compared the 1999-10-09 spectrum with that obtained on 1997-04-20 and found very little difference in the line intensities, again of the order of a few percent. This indicates that the basal, quiet-Sun mid-chromosphere is relatively stable with time, as one would expect. As in previous cases, the SUMER spectra were calibrated in wavelength by previous authors using reference lines from neutrals. It is unclear which lines were used, though. \cite{parenti_etal:2005a} refer to wavelengths and identifications for the carbon lines from \cite{kelly:1987}, however Kelly's compilation of these lines is based on the list by \cite{feldman_etal:1976_c_1}, which actually has, below 1150~\AA, the  wavelengths predicted by \cite{johansson:1965}. We have not carried out a careful wavelength calibration, but rather rely here on the \cite{curdt_etal:2001} calibration. We shall see below that there is generally good agreement between those 
wavelengths and our calculated values. 

In the SOHO SUMER spectra, the line profiles of the neutrals are mostly instrumental. \citet{chae_etal:1998} estimate an instrumental FWHM of 2.3 detector pixels, equivalent to 0.095\,\AA\ at 1500\,\AA. \citet{rao2022} estimate it to be closer to 0.11\,\AA, equivalent to a FWHM of 2.6 detector pixels. Using the \texttt{'CON\_WIDTH\_FUNCT\_3.PRO'} routine provided by the instrument team gives a corrected FWHM of 0.13\,\AA\ for detector B.
\citet{chae_etal:1998} report that there is little variation in the FWHM with wavelength. From the bin width in the observations at 1100\,\AA\ and 1500\,\AA, the FWHM changes by less than 4\% over this wavelength range. The thermal width of the lines is estimated to be 0.025\,\AA\ at 1460\,\AA\ for the chosen temperature of line formation, details of which are given below. 


\subsection{Modelling the Rydberg states}

\begin{table*}
\caption{ Sample of transition data provided electronically at ZENODO. $\lambda_{obs}$ and $\lambda_{calc}$ are the observed (if available, 
otherwise zero) and theoretical wavelengths (in \AA) of the transition, respectively; $f_{\rm L}$ and $f_{\rm V}$ are the absorption oscillator strengths in the length and velocity forms, respectively; $A_{ul}$ is the transition rate (in s$^{-1}$); $I_{up}$ is the ionisation energy (in Rydbergs) of the upper level to the respective parent; and $\epsilon_{ul}$ is the emissivity (in ergs s$^{-1}$), as defined in Sect.\;\ref{sec:comparison}, using an electron  temperature of 7000\,K and electron density of 6$\times$10$^{10}$\,cm$^{-3}$.}
\centering
\begin{tabular}{rrrrrllrr}
\hline
 $\lambda_{obs}$ & $\lambda_{calc}$ & $f_{\rm L}$ & $f_{\rm V}$ & $A_{ul}$  & Lower level & Upper level & $I_{up}$ & $\epsilon_{ul}$ \\
 \hline
 0.00 & 1102.557 & 9.358(-5) & 8.937(-5) & 1.712(+5) & 2s$^2$2p$^2$ $^3$P$_0$ & 2s$^2$2p($^2$P$_{1/2}^{\rm o}$) 30d $^2$[3/2]$^o_1$ & 1.11253(-3) & 3.355(-17) \\
 0.00 & 1102.558 & 5.455(-6) & 5.345(-6) & 9.977(+3) & 2s$^2$2p$^2$ $^3$P$_0$ & 2s$^2$2p($^2$P$_{1/2}^{\rm o}$) 31s $^2$[1/2]$^o_1$ & 1.11343(-3) & 1.955(-18) \\
 0.00 & 1102.617 & 2.550(-5) & 2.430(-5) & 4.663(+4) & 2s$^2$2p$^2$ $^3$P$_0$ & 2s$^2$2p($^2$P$_{3/2}^{\rm o}$) 24d $^2$[1/2]$^o_1$ & 1.73510(-3) & 9.266(-18) \\
 0.00 & 1102.618 & 6.034(-5) & 5.792(-5) & 1.103(+5) & 2s$^2$2p$^2$ $^3$P$_0$ & 2s$^2$2p($^2$P$_{3/2}^{\rm o}$) 24d $^2$[3/2]$^o_1$ & 1.73617(-3) & 2.192(-17) \\
 0.00 & 1102.626 & 8.681(-6) & 8.180(-6) & 1.588(+4) & 2s$^2$2p$^2$ $^3$P$_0$ & 2s$^2$2p($^2$P$_{3/2}^{\rm o}$) 25s $^2$[3/2]$^o_1$ & 1.74176(-3) & 3.156(-18) \\
 0.00 & 1102.662 & 1.104(-4) & 1.054(-4) & 2.019(+5) & 2s$^2$2p$^2$ $^3$P$_0$ & 2s$^2$2p($^2$P$_{1/2}^{\rm o}$) 29d $^2$[3/2]$^o_1$ & 1.19101(-3) & 3.963(-17) \\
\hline
\end{tabular}
\label{tab:datasample}
\end{table*}

To model emission from Rydberg states, use can be made of the fact that, at typical densities and temperatures in the mid-chromosphere, high-$n$ states should be close to local thermodynamic equilibrium (LTE). Their populations relative to the C$^+$ ground term from which they are recombining can be calculated using the Saha-Boltzmann equation. The number density, $N_u$ of a level $u$ relative to the number density, $N_p$, of its parent $p$ is

\begin{equation}
    \frac{N_u}{N_p} \;=\; \frac{g_u}{2g_p} \sqrt{\left( \frac{h^2}{2\pi m_e kT} \right)^3} \;{\rm exp}\left( \frac{I_{up}}{kT_e}\right) \; N_e ~,
    \label{eqn:saha}
\end{equation}

\noindent where $g$ is the statistical weight of a level, $m_e$ the electron mass, $I_{up}$ is the ionisation potential of the level relative to its parent, and $N_e$ the electron number density. 

There are two parents, C$^+(^2$P$_{1/2}^{\rm o})$ and C$^+(^2$P$_{3/2}^{\rm o})$ giving rise to bound Rydberg states. In the conditions under consideration, the high density ensures that their relative populations are determined by the Boltzmann equation. Their populations can be further assumed to be in the ratio of their statistical weights because they are so close in energy. We also assume that all of the C$^+$ population resides in the two lowest levels so that

\begin{equation}
      \frac{N_u}{N(C^+)} \;=\; \frac{g_u}{2g(C^+)} \sqrt{\left( \frac{h^2}{2\pi m_e kT} \right)^3} \;{\rm exp}\left( \frac{I_{up}}{kT_e}\right) \; N_e ~.
    \label{eqn:saha1}
\end{equation}

\noindent We define the emissivity here as the energy emitted per unit time per C$^+$ ion for each line emitted at wavelength $\lambda_{ul}$ in a transition from upper level $u$ to lower level $l$ as

\begin{equation}
    \epsilon_{ul} \;=\; \frac{hc}{\lambda_{ul}} \; \frac{N_u}{N(C^+)} \, A_{ul} ~,
    \label{eqn:emiss}
\end{equation}

\noindent where $A_{ul}$ is the transition probability. 
{ We do not calculate the number density of C$^+$, which is left as a free parameter, 
together with the carbon abundance. These free parameters
are included in the 
normalisation of the line intensities when we 
compare them with the solar spectra.

To estimate at which $n$ } the levels are likely to be in LTE, the helium CRM of \citet{delzanna_storey:2022} was run at $T_{\rm e}$=7\,000\,K and $N_{\rm e}$=$6\times 10^{10}$\,cm$^{-3}$. (These are the conditions at which \citealt{lin_etal:2017} state the 1355.85\,\AA\ \ion{C}{i} line forms.) The CRM predicts that states with $n \ge$11 have populations only 20\% lower than the LTE values. Those with $n \ge$15 have departure coefficients of 0.9 and above. The published data from the hydrogen CRM of \citet{Hummer1987} are broadly in agreement, predicting that at 7\,500\,K the H populations are within 10\% of LTE values at $n=9$ at an electron density of $10^{11}$\,cm$^{-3}$  and at $n=11$ at a density of $10^{10}$\,cm$^{-3}$. LTE values are reached at slightly lower values of $n$ for H compared to He because collision rates between the degenerate levels of H are faster than those between the non-degenerate levels of He.

While there are inevitably differences between the H and He CRMs and the present model, it highlights that for states with $n \ge$11 LTE is a good approximation to calculate the line emissivities within a 20\% error, assuming these lines are formed at those densities and temperatures. There is a further potential  uncertainty in the populations for the lower levels in LTE relative to the highest levels, arising from the exponential term in equation\;(\ref{eqn:saha1}), depending on the temperature where the lines form. However, the exponential varies by only 15\% for the $n$=11 levels relative to the highest levels in the temperature range { between}  5\,000\,K and 12\,000\,K. So, it is reasonable to compare relative line intensities of the transitions, and the emission measure can be assumed, to a first approximation, to be the same for all levels in LTE. Relative line intensities are assessed by normalising the synthetic spectrum to one or two observed lines emitted from levels with $n\approx 20$ in each series, and then comparing the agreement between observation and theory along the series. Only statistical weights, ionisation energies and $A$-values are required for the comparison.

In the first instance, we ignore the potential effect of optical depth on the lines and assume that the line photons are emitted from a region of fixed temperature and density and escape freely from the medium. Comparing the predicted and observed intensities of lines from a common upper level to different lower levels offers information about the degree to which radiative transfer affects the emergent intensities between the different series. This assumption will be discussed further below.  


Based on the observations, the line shapes appear to be Gaussian. The synthetic spectrum for each transition was calculated from

\begin{equation}
    I_{ul} \;=\; \frac{1}{\sigma\sqrt{2\pi}}\, {\rm exp} \left( -\frac{(\lambda - \lambda_{ul})^2}{2\sigma^2}\right) \, \epsilon_{ul} ~,
\end{equation}

\noindent where $\rm FWHM={\sigma} \sqrt{8\,{\rm ln}2}$ and FWHM is the observed full-width half maximum of the Gaussian (0.13\,\AA\ for these observations).

Separation of the lines from the continuum in the observations poses some problems. The $R$-matrix code allows the calculation of photoionisation cross sections for the \ion{C}{i} ground configuration from which a synthetic continuum was derived. This was problematic for various reasons, not least because it is obvious that continua from other elements are present in the observations. An alternative approach of using the continuum from the radiative transfer, hydrostatic calculation of \citet{fontenla_etal:2014} was attempted, but this did not improve the comparison. Finally, the continuum was subtracted from the observations by taking the minimum intensity at every 1\,\AA\ interval (for the $^1$S transitions) or 2\,\AA\ interval (for the other series) and subtracting that value from the observed intensities in that wavelength interval. Despite the limitations of this approach, it allowed a meaningful comparison.

\begin{figure*}
\centering
\includegraphics{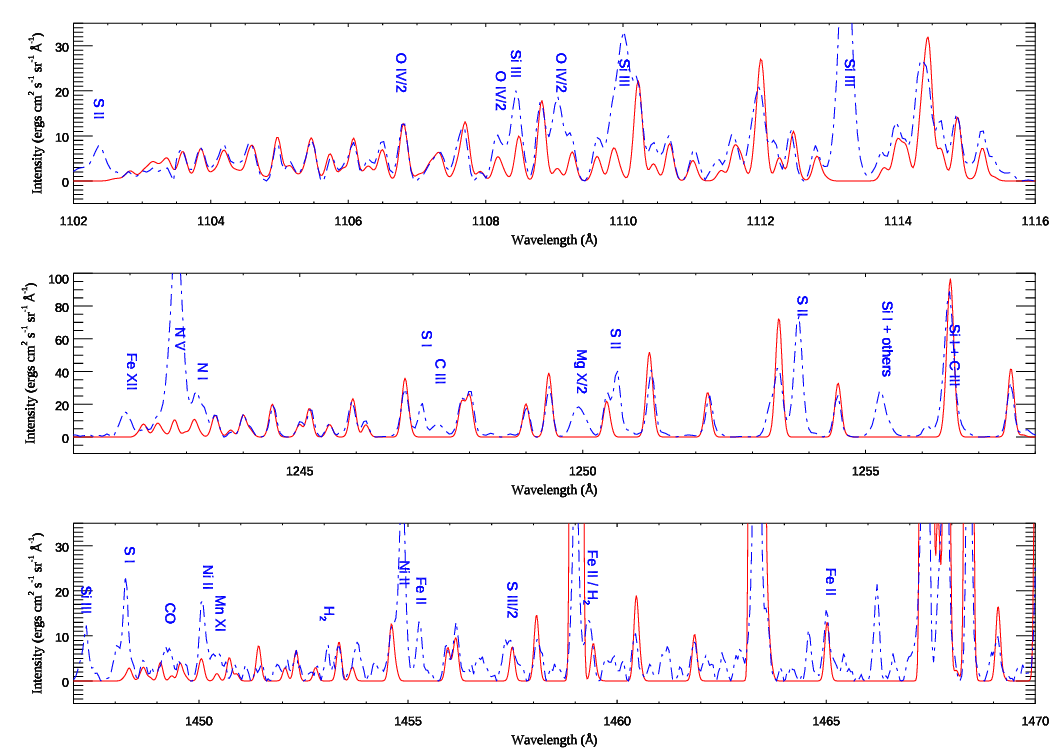}
\caption{Comparison of synthetic spectrum with SUMER observations: red solid line - synthetic spectrum; blue dash-dotted - \citeauthor{curdt_etal:2001}. Main lines in the solar spectrum emitted by other ions are marked in blue at their observed wavelength. A different normalisation of the synthetic spectrum is used in each subplot. See text for more details.}
\label{fig:spectrum_indep}
\end{figure*}


\subsection{Comparison with observations - optically thin case}

The comparison with the SUMER spectrum is illustrated in Fig.\;\ref{fig:spectrum_indep}. Each subplot shows the spectrum from the highest levels at $n=30$ down to $n=10$, where the levels should begin to depart from LTE. It is clear from the comparison that the agreement obtained using the new atomic data is excellent, both in outline and in detail. Looking first at the transitions to the 2s$^2$\,2p$^2\;^3$P term close to 1100\,\AA, the synthetic spectrum captures very well the details in the self-blends found at 1106.3\,\AA, 1107.2\,\AA\ and 1107.9\,\AA, for example. Subtraction of the continuum is more problematic in this wavelength range because the transitions from each set of $n$d/($n$+1)s levels to the three $^3$P$_J$ levels are closely spaced. The strong line at 1112.0\,\AA\ (from an 11d level) may show the first signs of departure from LTE in the observations, although it should be remembered that there is approximately 15\% uncertainty in their populations relative to the highest levels depending on the temperature at which the lines form. The weaker lines observed in the range 1113-1116\,\AA, from the 10d/11s levels, appear to show the more likely signs that the levels may be below LTE populations, but optical depth effects and subtracting the continuum cannot be ruled out from causing this, as well. Overall, in this wavelength range the comparison is not affected significantly by blends from other ions.

In solar observations, the \ion{N}{v} line at 1242.8\,\AA\ obscures decays to 2s$^2$\,2p$^2\;^1$D$_2$ from the highest levels in the calculation. (The series limit is 1240.27\,\AA\ and the lines are shown in the middle subplot of Fig.\;\ref{fig:spectrum_indep}.) From wavelengths longer than 1243.5\,\AA\ (from both the 2s$^2\,2$p\,$^2$P$_{1/2}^{\rm o}$\,23d and 2s$^2\,2$p\,$^2$P$_{3/2}^{\rm o}$\,20d $J$=3 levels) all of the neutral carbon lines are clearly visible, plus there is little difficulty in subtracting the continuum in this region. For this series the agreement is also excellent, with much of the detail matching observations very well. While there are some small over-predictions in the relative synthetic intensities beginning at 1246.9\,\AA\ and 1249.4\,\AA\ (from the 15d/16s and 13d/14s levels respectively), the theory is obviously higher for the stronger line at 1253.5\,\AA\ (from an 11d $J$=3 level). Again, as with the $^3$P series, this over-prediction occurs more for the stronger lines, perhaps indicating that it relates more to optical depth effects. The effect cannot be seen for the 1256.5\,\AA\ line because it is blended with a \ion{Si}{i} line and \ion{C}{iii} lines. Because of blends with $^3$P transitions in the wavelength range 1260-1262\,\AA, systematic differences between theory and observations cannot be seen until $n$=8 (around 1266\,\AA) in this series. For the $^1$D transitions shown in Fig.\;\ref{fig:spectrum_indep}, a different normalisation is used to fit the synthetic spectrum to the observations compared to the normalisation required for the transitions to the $^3$P states. The implications of this will be discussed in the next section.

In the wavelength range of the decays to 2s$^2$\,2p$^2\;^1$S$_0$ there are very many weak lines present in the observations, as shown in the lower subplot of Fig.\;\ref{fig:spectrum_indep}. The series limit (at 1445.67\,\AA) is somewhat obscured by lines from \ion{Si}{iii} and \ion{S}{i}. The remainder of the neutral carbon lines in this range are clear as a whole and the theory again matches the observations very well. More small over-predictions are seen, at 1458.0\,\AA\ and 1460.4\,\AA, but the systematic differences begin at 1469\,\AA, for decays from $n$=10 and lower. The lines at 1459.1\,\AA, 1463.3\,\AA\ and 1467.1-1468.6\,\AA\ are decays from the 3d/4s levels to the $^1$D$_2$ level, and are not relevant for the present comparison.

\subsection{Estimate of optical depth effects}

While the theoretical line intensities within each series can be matched to observations with a single normalisation, the normalisation required for each series is different. If the normalisation for the $^1$S series is applied to the $^3$P series, the observations are clearly weaker than expected from theory. Applying the same normalisation to the $^1$D series also shows that those observations are weaker than theory, but not by as much as the $^3$P series. This is the kind of behaviour expected if there are optical depth effects in the lines: there should be more absorption from the more highly populated $^3$P levels, for example, resulting in photons emitted in the $^3$P series originating higher in the atmosphere than those in the $^1$D and $^1$S series.  The effect is also consistent with the discussion in the previous section, that in some cases the stronger lines appear to be weaker than theory even when the levels are expected to be in LTE.

An estimation of the optical depth effects can be made by comparing the relative intensities of decays from common upper levels to different lower levels. This approach reduces the  uncertainty in the population of the upper levels. The wavelength range 1113.5--1115.5 \AA\ includes all the transitions from the 2s$^2$\,2p\,($^2{\rm P}^o$)\,10d and 11s states to the ground $^3$P$_J$ levels, while the range 1256.0--1258.0 \AA\ includes the transitions from the same upper states to the $^1$D$_2$ level.  \citet{VALIII} presented a chromospheric model in which they calculated and tabulated temperatures and particle number densities as a function of height in the atmosphere, measured from unit optical depth in the continuum at 500~nm, $\tau_{500}$. The benefit of their calculation is that they included number densities for neutral and once ionized carbon. In their model C, they tabulate this information for the quiet Sun, for each of the fifty-two layers of their plane-parallel model atmosphere, making it possible to calculate the line emissivities and optical thicknesses of each layer. We use an escape probability formalism (\cite{H&S1992} and references therein) to calculate the total energy escaping from the atmosphere as a function of wavelength in the two wavelength ranges above. 

\begin{figure*}
\centering
\includegraphics[width=1.5\columnwidth, angle=-90]{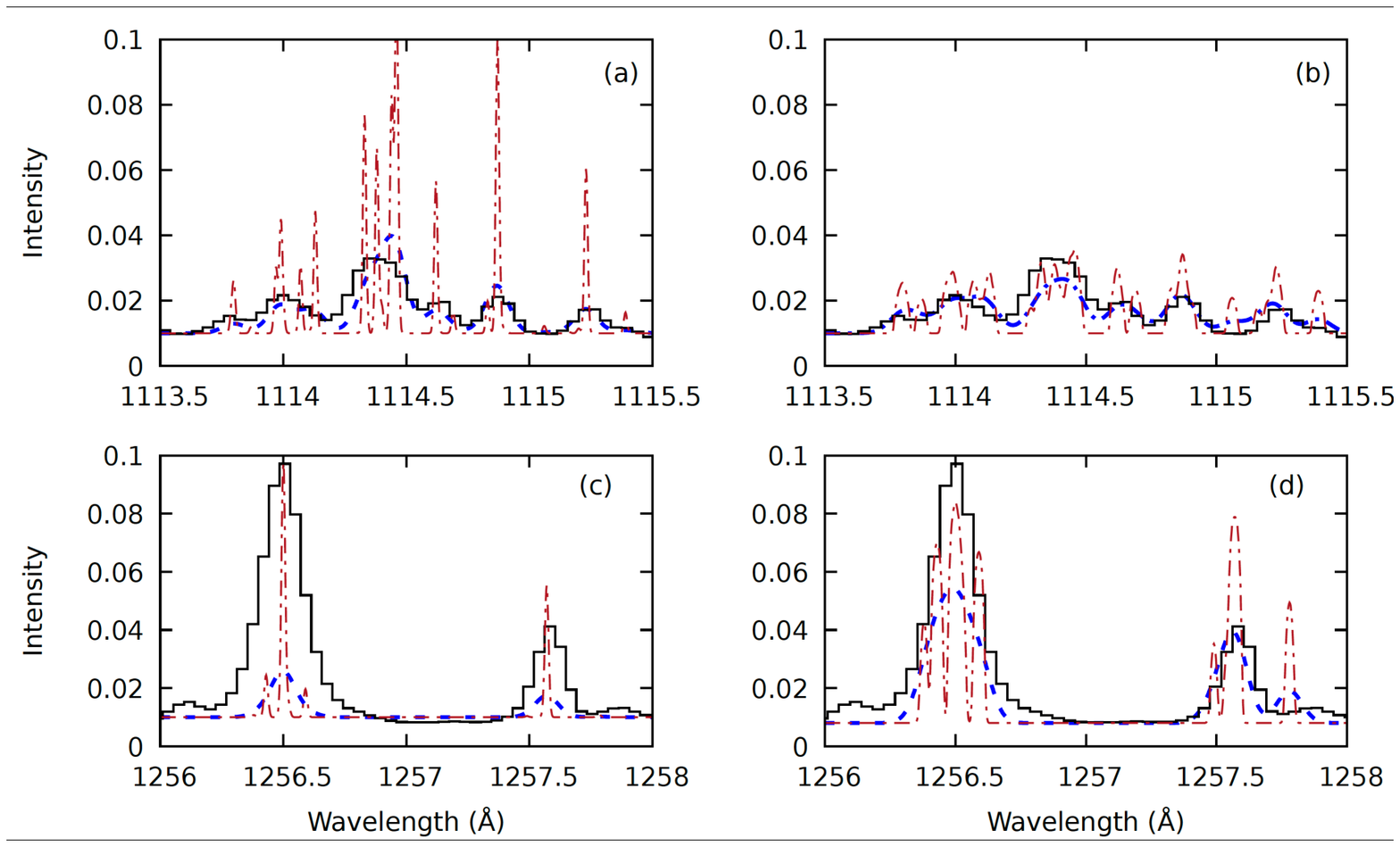}
\caption{Comparison of synthetic spectra with SUMER observations for lines from the $n=10$d and 11s levels: red dot-dash line - synthetic emergent spectrum; blue dashed line - emergent spectrum convolved with an instrumental profile; black solid line - observations  \citeauthor{curdt_etal:2001}. The two left hand panels are the results of an optically thin model, while the two right hand panels are from an optically thick model. The two upper panels are transitions decaying to the ground $^3$P$_J$ levels while the lower panels are transitions decaying to the $^1$D$_2$.  See the text for further explanation.}
\label{fig:spectrum_n10}
\end{figure*}

 Assuming a semi-infinite, plane parallel atmosphere comprising $n$(=52) layers of physical thickness $\Delta s$, the intensity of radiation  escaping the atmosphere normal to the surface at wavelength $\lambda$ is 

\begin{equation} 
  I(\lambda) = \Sigma_{m=1}^n I_m(\lambda) = \Sigma_{m=1}^n\, \frac{1}{4\pi}\,\epsilon_m(\lambda)\, {\rm e}^{-\tau_m(\lambda)}\, \Delta{\rm s}_m ,
\end{equation}
where $I_m(\lambda)$ is the intensity emerging from layer $m$ and $\tau_m(\lambda)$ is the total optical depth to the surface ($m=1$), given by
\begin{equation}
    \tau_m(\lambda) = \Sigma_{k=1}^m T_k(\lambda)
    \end{equation}
where $T_k(\lambda)$ is the optical thickness of layer $k$ given by
\begin{equation}
   T_k(\lambda) = \frac{\pi\, {\rm e}^2}{{\rm m\, c}}\, \Delta s_k\, \Sigma_{\rm lines} f_{lu}\, N_l\, \phi(\lambda_{ul},\lambda)
\end{equation}
and
\begin{equation}
    \epsilon_m(\lambda) = \Sigma_{\rm lines}\, N_u\, A_{ul}\, {\rm h}\nu_{ul}\, \phi(\lambda_{ul},\lambda) 
\end{equation}
is the emissivity from layer $k$ at wavelength $\lambda$.  The sums over lines add the contributions to the emissivity and optical thickness from all nearby lines between an upper state $u$, number density $N_u$, and lower state $l$, number density $N_l$, with the corresponding oscillator strength $f_{lu}$, transition probability $A_{ul}$ and central wavelength $\lambda_{ul}$. The line profile function $\phi$ is assumed to be Gaussian with a Doppler width corresponding to the local electron temperature. Once the emergent intensity is calculated at each wavelength it can then be convolved with an instrumental profile, again assumed Gaussian, with a width of 0.11\AA . Only line emission and absorption are included, continuum processes are neglected. 

In Figure~\ref{fig:spectrum_n10} we compare the results of an optically thin calculation with the escape probability model to assess the effects of optical depth. The dot-dashed lines show the calculated emergent intensity, the dashed line shows the convolved spectrum and the solid line is the SUMER observations. Panel (a) of Figure~\ref{fig:spectrum_n10} shows a model in which all optical depths were set to zero, so that all photons escape, and the convolved spectrum has been scaled to match the observations. These are all lines from the 10d, 11s group of Rydberg states decaying to the $^3$P$_J$ ground levels. Panel (c) of Figure~\ref{fig:spectrum_n10} shows the predicted intensities of the same group of upper states decaying to the ground configuration $^1$D$_2$ level, using the same scaling as in panel (a). The optically thin model under predicts the intensities of lines to the $^1$D$_2$ level by a factor of four to six. In the chromosphere, the population of the $^1$D$_2$ level is much less than that of the $^3$P$_J$ levels and the lines ending on that state are less optically thick and are  being formed deeper in the atmosphere and over a longer path length, which leads to a greater intensity. Panels (b) and (d) show the same comparison for the optically thick model. Again the convolved spectrum has been scaled to the observations in the 1113.5 to 1115.5\,\AA\ range but the agreement is now much improved for the decays to the $^1$D$_2$ level. The lines at 1115.23\,\AA\ and 1257.57\,\AA\ originate from the same upper level and at line center the optical depth to the surface is of order unity for the 1115.23\,\AA\ line at a height of 1990\,km where the model temperature is 7160\,K, while for the 1257.57\,\AA\ line  this occurs lower at 1065\,km where the temperature is 6040\,K. For comparison, the \ion{C}{i} continuum optical depth at 1115.23\,\AA\ reaches unity  much deeper in the atmosphere, near the temperature minimum at a height of 450\,km.

There are some features of the optically thick model spectra that merit comment. The profiles of the lines in panels (b) and (d) are significantly broader than the Doppler profiles  seen in the optically thin case in panels (a) and (c). Also some very weak lines, for example the line at 1257.8\,\AA, are strongly amplified and predicted to be stronger than observed. This is probably an artifact of the calculation, in that using a Gaussian profile means that the optical depth can become vanishingly small in the wings of the line. This means that photons can escape from very deep in the atmosphere from regions where the emissivity is very high. In such regions, the continuum opacity, which we neglect, should be taken into account and would prevent the escape of such photons.  The red wing of Ly-$\alpha$  also contributes strongly to the opacity in the 1256.0--1258.0\,\AA\ region \citep{fontenla_etal:2014}. The same argument applies to intrinsically very weak lines. 

We mentioned previously that different scalings were needed to match theory to observation when comparing lines decaying to the $^3$P$_J$, $^1$D$_2$ or $^1$S$_0$ levels. The results of the optically thick model indicate that these differences are due to the effects of optical depth rather than other causes, such as the atomic data, for example.  We note that essentially the same results are obtained by solving the equation of radiative transfer  in one dimension, including only \ion{C}{i} line emission and absorption and ignoring continuum processes.






\subsection{New identifications}

\begin{table}
\caption{List of previously unidentified transitions contributing to the observed intensity of solar lines. An asterisk indicates that the transition is also not included in the NIST database. $\lambda$ is the observed wavelength (in \AA) of the transition if available, otherwise the calculated wavelength; $\epsilon_{ul}$ is the emissivity (in ergs s$^{-1}$) defined in the text.}
\centering
\setlength\tabcolsep{4pt}
\begin{tabular}{llll}
\hline
 $\lambda$ & Lower level & Upper level & $\epsilon_{ul}$ \\
\hline
1110.44 & 2s$^2$\,2p$^2$\;$^3$P$_0$ & 2s$^2$\,2p\,($^2$P$^{\rm o}_{1/2}$)\,12d\;$^2$[3/2]$^{\rm o}_1$ & 6.006(-16) \\
1110.68 & 2s$^2$\,2p$^2$\;$^3$P$_1$ & 2s$^2$\,2p\,($^2$P$^{\rm o}_{1/2}$)\,12d\;$^2$[3/2]$^{\rm o}_2$ & 1.264(-15) \\
1128.07 & 2s$^2$\,2p$^2$\;$^3$P$_0$ &  2s$^2$\,2p\,($^2$P$^{\rm o}_{3/2}$)\,7d\;$^2$[1/2]$^{\rm o}_1$ & 6.437(-16) \\
1128.17 & 2s$^2$\,2p$^2$\;$^3$P$_0$ &  2s$^2$\,2p\,($^2$P$^{\rm o}_{3/2}$)\,7d\;$^2$[3/2]$^{\rm o}_1$ & 5.528(-16) \\
1128.26 & 2s$^2$\,2p$^2$\;$^3$P$_1$ &  2s$^2$\,2p\,($^2$P$^{\rm o}_{3/2}$)\,7d\;$^2$[1/2]$^{\rm o}_0$ & 1.500(-15) \\
1128.28 & 2s$^2$\,2p$^2$\;$^3$P$_1$ &  2s$^2$\,2p\,($^2$P$^{\rm o}_{3/2}$)\,7d\;$^2$[1/2]$^{\rm o}_1$ & 1.429(-15) \\
1128.62 & 2s$^2$\,2p$^2$\;$^3$P$_2$ &  2s$^2$\,2p\,($^2$P$^{\rm o}_{3/2}$)\,7d\;$^2$[1/2]$^{\rm o}_1$ & 1.646(-15) \\
1128.69 & 2s$^2$\,2p$^2$\;$^3$P$_1$ &  2s$^2$\,2p\,($^2$P$^{\rm o}_{3/2}$)\,8s\;$^2$[3/2]$^{\rm o}_1$ & 1.611(-15) \\
1128.72 & 2s$^2$\,2p$^2$\;$^3$P$_2$ &  2s$^2$\,2p\,($^2$P$^{\rm o}_{3/2}$)\,7d\;$^2$[3/2]$^{\rm o}_1$ & 9.216(-18) \\
1128.72 & 2s$^2$\,2p$^2$\;$^3$P$_2$ &  2s$^2$\,2p\,($^2$P$^{\rm o}_{3/2}$)\,7d\;$^2$[3/2]$^{\rm o}_2$ & 4.513(-15) \\
1128.82 & 2s$^2$\,2p$^2$\;$^3$P$_1$ &  2s$^2$\,2p\,($^2$P$^{\rm o}_{3/2}$)\,7d\;$^2$[5/2]$^{\rm o}_2$ & 7.572(-15) \\
1128.90 & 2s$^2$\,2p$^2$\;$^3$P$_2$ &  2s$^2$\,2p\,($^2$P$^{\rm o}_{3/2}$)\,7d\;$^2$[7/2]$^{\rm o}_3$ & 2.756(-15) \\
1129.03 & 2s$^2$\,2p$^2$\;$^3$P$_2$ &  2s$^2$\,2p\,($^2$P$^{\rm o}_{3/2}$)\,8s\;$^2$[3/2]$^{\rm o}_1$ & 5.770(-16) \\
1129.08 & 2s$^2$\,2p$^2$\;$^3$P$_1$ &  2s$^2$\,2p\,($^2$P$^{\rm o}_{3/2}$)\,8s\;$^2$[3/2]$^{\rm o}_2$ & 7.876(-16) \\
1129.20 & 2s$^2$\,2p$^2$\;$^3$P$_0$ &  2s$^2$\,2p\,($^2$P$^{\rm o}_{1/2}$)\,7d\;$^2$[3/2]$^{\rm o}_1$ & 4.058(-15) \\
1129.32 & 2s$^2$\,2p$^2$\;$^3$P$_0$ &  2s$^2$\,2p\,($^2$P$^{\rm o}_{1/2}$)\,8s\;$^2$[1/2]$^{\rm o}_1$ & 9.748(-16) \\
1129.42 & 2s$^2$\,2p$^2$\;$^3$P$_2$ &  2s$^2$\,2p\,($^2$P$^{\rm o}_{3/2}$)\,8s\;$^2$[3/2]$^{\rm o}_2$ & 6.197(-15) \\
1139.30 & 2s$^2$\,2p$^2$\;$^3$P$_1$ &  2s$^2$\,2p\,7s\;$^1$P$^{\rm o}_1$ & 2.706(-15) \\
1139.43 & 2s$^2$\,2p$^2$\;$^3$P$_2$ &  2s$^2$\,2p\,6d\;$^1$F$^{\rm o}_3$ & 3.494(-15) \\
1139.51 & 2s$^2$\,2p$^2$\;$^3$P$_1$ &  2s$^2$\,2p\,6d\;$^3$D$^{\rm o}_2$ & 3.764(-15) \\
1139.79 & 2s$^2$\,2p$^2$\;$^3$P$_0$ &  2s$^2$\,2p\,6d\;$^3$D$^{\rm o}_1$ & 7.322(-15) \\
1193.39 & 2s$^2$\,2p$^2$\;$^3$P$_2$ &  2s$^2$\,2p\,4d\;$^3$D$^{\rm o}_2$ & 4.282(-14) \\
1194.00 & 2s$^2$\,2p$^2$\;$^3$P$_0$ &  2s$^2$\,2p\,5s\;$^3$P$^{\rm o}_1$ & 1.579(-14) \\
1194.61 & 2s$^2$\,2p$^2$\;$^3$P$_2$ &  2s$^2$\,2p\,5s\;$^3$P$^{\rm o}_1$ & 1.465(-14) \\
1448.33 & 2s$^2$\,2p$^2$\;$^1$S$_0$ & 2s$^2$\,2p\,($^2$P$^{\rm o}_{3/2}$)\,24d\;$^2$[3/2]$^{\rm o}_1$ & 1.838(-17) \\
1448.60* & 2s$^2$\,2p$^2$\;$^1$S$_0$ & 2s$^2$\,2p\,($^2$P$^{\rm o}_{1/2}$)\,28d\;$^2$[3/2]$^{\rm o}_1$ & 9.385(-18) \\
1448.68* & 2s$^2$\,2p$^2$\;$^1$S$_0$ & 2s$^2$\,2p\,($^2$P$^{\rm o}_{3/2}$)\,23d\;$^2$[3/2]$^{\rm o}_1$ & 1.955(-17) \\
1448.82* & 2s$^2$\,2p$^2$\;$^1$S$_0$ & 2s$^2$\,2p\,($^2$P$^{\rm o}_{1/2}$)\,27d\;$^2$[3/2]$^{\rm o}_1$ & 8.127(-18) \\
1449.07* & 2s$^2$\,2p$^2$\;$^1$S$_0$ & 2s$^2$\,2p\,($^2$P$^{\rm o}_{1/2}$)\,26d\;$^2$[3/2]$^{\rm o}_1$ & 2.052(-17) \\
1449.09* & 2s$^2$\,2p$^2$\;$^1$S$_0$ & 2s$^2$\,2p\,($^2$P$^{\rm o}_{3/2}$)\,22d\;$^2$[3/2]$^{\rm o}_1$ & 1.288(-17) \\
1449.35* & 2s$^2$\,2p$^2$\;$^1$S$_0$ & 2s$^2$\,2p\,($^2$P$^{\rm o}_{1/2}$)\,25d\;$^2$[3/2]$^{\rm o}_1$ & 1.153(-17) \\
1449.55* & 2s$^2$\,2p$^2$\;$^1$S$_0$ & 2s$^2$\,2p\,($^2$P$^{\rm o}_{3/2}$)\,21d\;$^2$[3/2]$^{\rm o}_1$ & 2.836(-17) \\
1449.67* & 2s$^2$\,2p$^2$\;$^1$S$_0$ & 2s$^2$\,2p\,($^2$P$^{\rm o}_{1/2}$)\,24d\;$^2$[3/2]$^{\rm o}_1$ & 1.033(-17) \\
1450.43* & 2s$^2$\,2p$^2$\;$^1$S$_0$ & 2s$^2$\,2p\,($^2$P$^{\rm o}_{1/2}$)\,22d\;$^2$[3/2]$^{\rm o}_1$ & 1.690(-17) \\
1450.71* & 2s$^2$\,2p$^2$\;$^1$S$_0$ & 2s$^2$\,2p\,($^2$P$^{\rm o}_{3/2}$)\,19d\;$^2$[1/2]$^{\rm o}_1$ & 1.044(-17) \\
1450.73* & 2s$^2$\,2p$^2$\;$^1$S$_0$ & 2s$^2$\,2p\,($^2$P$^{\rm o}_{3/2}$)\,19d\;$^2$[3/2]$^{\rm o}_1$ & 3.846(-17) \\
1450.90* & 2s$^2$\,2p$^2$\;$^1$S$_0$ & 2s$^2$\,2p\,($^2$P$^{\rm o}_{1/2}$)\,21d\;$^2$[3/2]$^{\rm o}_1$ & 1.633(-17) \\
1452.06* & 2s$^2$\,2p$^2$\;$^1$S$_0$ & 2s$^2$\,2p\,($^2$P$^{\rm o}_{1/2}$)\,19d\;$^2$[3/2]$^{\rm o}_1$ & 2.932(-17) \\
1452.80* & 2s$^2$\,2p$^2$\;$^1$S$_0$ & 2s$^2$\,2p\,($^2$P$^{\rm o}_{1/2}$)\,18d\;$^2$[3/2]$^{\rm o}_1$ & 3.036(-17) \\
1453.67* & 2s$^2$\,2p$^2$\;$^1$S$_0$ & 2s$^2$\,2p\,($^2$P$^{\rm o}_{1/2}$)\,17d\;$^2$[3/2]$^{\rm o}_1$ & 3.182(-17) \\
1454.59* & 2s$^2$\,2p$^2$\;$^1$S$_0$ & 2s$^2$\,2p\,($^2$P$^{\rm o}_{3/2}$)\,15d\;$^2$[1/2]$^{\rm o}_1$ & 2.126(-17) \\
1454.64* & 2s$^2$\,2p$^2$\;$^1$S$_0$ & 2s$^2$\,2p\,($^2$P$^{\rm o}_{3/2}$)\,16s\;$^2$[3/2]$^{\rm o}_1$ & 2.092(-17) \\
1454.71* & 2s$^2$\,2p$^2$\;$^1$S$_0$ & 2s$^2$\,2p\,($^2$P$^{\rm o}_{1/2}$)\,16d\;$^2$[3/2]$^{\rm o}_1$ & 2.278(-17) \\
1455.95* & 2s$^2$\,2p$^2$\;$^1$S$_0$ & 2s$^2$\,2p\,($^2$P$^{\rm o}_{1/2}$)\,15d\;$^2$[3/2]$^{\rm o}_1$ & 7.870(-17) \\
 \hline
 \hline
\end{tabular}
\label{tab:newidents}
\end{table}

There are many more lines unidentified in the 1450\,\AA\ wavelength range of the solar spectrum compared to the other wavelength regions illustrated in Fig.\;\ref{fig:spectrum_indep}. It is also not certain whether the identifications from \citet{curdt_etal:2001}, \citet{sandlin_etal:1986} and \citet{parenti_etal:2005a} shown in Fig.\;\ref{fig:spectrum_indep}, such as the \ion{Fe}{ii} 1241.9\,\AA\ and CO 1449.2\,\AA\ lines, are the only contributors to the observed intensity of each line. 

Out of the new identifications provided by the present calculation, the majority involve decays to the 2s$^2$\,2p$^2\;^1$S$_0$ level, which is understandable given the few atomic calculations involving this level. Table\;\ref{tab:newidents} identifies lines which are not listed by the solar observations referenced in the previous paragraph, but which should contribute an observable amount in the quiet Sun spectrum. In addition, some of these lines are not present in the NIST database; they have been highlighted by an asterisk in the table. 

Since SUMER requires known lines in each 43\,\AA\ window, the new identifications could assist with calibration of the instrument, especially in the 1448-1610\,\AA\ region. For instance, the SUMER observations at most wavelengths show good agreement with the present calculation, but there is a discrepancy around 1512\,\AA. The SUMER observations for carbon peak at 1511.83\,\AA, while the theoretical wavelength is 1511.79\,\AA\ and the experimental wavelength from NIST is 1511.91\AA. \citet{sandlin_etal:1986} indicates an unknown line at 1511.84\,\AA, a carbon line of slightly higher intensity at 1511.91\,\AA\ and a weaker \ion{Fe}{ii} line at 1512.06\,\AA. The first two lines are blended in SUMER, but SUMER has the \ion{Fe}{ii} line peaking at 1512.00\,\AA, a shift of 0.06\,\AA\ from HRTS. This is similar to the difference between the SUMER and NIST wavelengths for the carbon line.

To illustrate how these new identifications affect interpretation of observations, the \ion{Si}{iv} line at 1128.35\,\AA\ can be considered. \citet{dufresne2023piobs} recently assessed how new atomic models for the solar transition region altered emission from \ion{Si}{iv}. Predictions for the \ion{Si}{iv} 1128.35\,\AA\ line were further from observations than predictions for the resonance lines. Because the upper level which emits the 1128.35\,\AA\ line is much higher in energy, such a discrepancy could indicate the influence of time dependent ionisation or non-Maxwellian electron distributions enhancing the emission from highly excited levels \citep[see][for example]{pietarila2004}. However, the \ion{Si}{iv} line at 818.15\,\AA\ is emitted from an upper level higher in energy than the 1128.35\,\AA\ line, and yet its predicted to observed intensity ratio agrees with the resonance lines. The 818.15\,\AA\ is a weak line and there may be greater uncertainty in its intensity. However, the present work highlights that \ion{C}{i} lines at 1128.26\,\AA\ and 1128.28\,\AA\ appear to be contributing to the observed intensity of the \ion{Si}{iv} 1128.35\,\AA\ line. This may account for at least part of the discrepancy in the 1128.35\,\AA\ line compared to the resonance lines. More detailed modelling of \ion{C}{i} emission would be required to determine to what extent it would bring predictions for the 1128.35\,\AA\ line into agreement with the other \ion{Si}{iv} lines.

The present work could also help in other areas of astrophysics, such as the interpretation of absorption lines for interstellar abundances mentioned in Sect.\;\ref{sec:intro}. Molecular HD is used as a probe for chemical evolution and nucleosynthesis studies. A weak HD line at 1105.83\,\AA\ is useful in the analysis because it is optically thin and unsaturated compared to the stronger HD lines at shorter wavelengths \citep{snow2008}. The line, however, is blended with the \ion{C}{i} line at 1105.72\,\AA. Analyses have often relied on the compilation of empirically determined oscillator strengths by \citet{morton1978}. \citet{snow2008} re-assessed the oscillator strength for the 1105.72\,\AA\ line by comparing the \citet{morton1978} data for lines at lower $n$ with the oscillator strengths of \citet{oleg:2002}. A downward revision by a factor of two did not significantly affect column densities and the subsequent analysis of the HD line, but \citet{snow2008} assume in their model that the 1105.72\,\AA\ line is a single transition from the lower $^3$P$_1$ level. There are 14 transitions in the range 1105.72-1105.77\,\AA, of which seven could contribute to the observed intensity. The present work shows that all three $^3$P$_J$ levels contribute to the feature, with the strongest line being from the $^3$P$_0$ level. \citet{snow2008} also note a further source of systematic error due to the absorption feature at 1105.92\,\AA, which was still unidentified at the time. The present work shows that neutral carbon contributes at this wavelength; the main contribution is from the 2s$^2$\,2p$^2$\;$^3$P$_1$ - 2s$^2$\,2p\,($^2$P$^o_{1/2}$)\,17d\;$^2$[3/2]$^o_2$ transition.

\section{Conclusions}
\label{sec:conclusions}
The frozen cores approximation and the Breit–Pauli $R$-matrix method are confirmed as powerful tools to calculate accurate energies and radiative data for neutral carbon. The accuracy of the energies of the Rydberg states is comparable to experimental values. Given the relatively large discrepancies in wavelengths found in the literature, the theoretical energies could be used to improve wavelength calibrations.

Using our previous collisional-radiative models for neutrals, we have estimated that for typical chromospheric conditions where neutral carbon is formed in the solar atmosphere, Rydberg states should be in LTE. The comparison with observations confirms this, showing excellent agreement in relative intensities for each series decaying to the three terms in the ground configuration. All levels with $n\ge11$ appear to be in LTE, with only a few differences in some strong lines, which are likely to be caused by optical depth effects.  This means that the Rydberg lines can be used as a diagnostic tool to probe the chromosphere without the need for a full collisional-radiative population model. 

{ Oscillator strengths are very close to earlier calculations for $n\le10$, usually agreeing withing 10\%, not just with similar $R$-matrix calculations but also those using other methods. The only discrepancies appear to be with the calculation of \citet{haris_kramida:2017ApJS..233...16H} above $n\ge9$ using the \citet{cowan:1981} code, which would not be expected to perform so well for Rydberg states.  In the present calculation, oscillator strengths in the length and velocity formulations agree within 10\% and, with the exception of one series, this good agreement persists up to the highest $n=30$. 
This gives us confidence that the uncertainty in our calculated oscillator strengths for the Rydberg states is of this order, which is further confirmed by the 
comparisons with the observations.}

Our relatively simple, escape probability, 1-D  model clearly indicates that lines of different series (and the continua) form at different depths in the chromosphere. Hence, their diagnostic must take that into account. The escape probability model was able to account for much of the absorption noted in observations for lines decaying to the $^3$P$_J$ levels, relative to the intensities of the lines in the $^1$D$_2$ series of decays.

The accuracy in theoretical wavelengths of the Rydberg states in the present calculation is also sufficient for them to be used as standards when measuring Doppler shifts. The accuracy of the wavelengths and intensities was also sufficient to enable us to identify for the first time a large number of transitions, 19 of which are not in the NIST line list, and to correct a few previous identifications. It also enabled us to explain some of the discrepancies for the 1128.35\,\AA\ \ion{Si}{iv} line compared to the resonance lines from the same ion. The present data are useful for analyses in many other areas of astrophysics, such as the interstellar abundances scenario highlighted in the Introduction.

The present work is the first stage in a proposed plan to build a collisional-radiative model for neutral carbon which includes the atomic rates from detailed calculations. This will be required to not only improve the modelling of the stronger spectral lines arising from lower states in the solar atmosphere, but also to investigate in more detail the radiative transfer effects that have been highlighted in this work. All together, these tools are expected to produce new diagnostics for those exploring the poorly-understood solar chromosphere.


\section*{Acknowledgments}
We acknowledge support from STFC (UK) via the consolidated grants 
to the atomic astrophysics group (AAG) at DAMTP, University of Cambridge (ST/T000481/1 and ST/X001059/1).
We thank the reviewer for the detailed comments.

\section*{Data Availability}

{ 
The full set of atomic data are available at ZENODO
(https://doi.org/10.5281/zenodo.8225754)

}

\bibliographystyle{mn2e}

\bibliography{paper}


\end{document}